\documentclass[aps,preprintnumbers,showpacs,twocolumn,superscriptaddress,nofootinbib]{revtex4}

\usepackage{amsmath}
\usepackage{amssymb}
\usepackage{color}
\usepackage{dcolumn}
\usepackage{graphicx}
\usepackage{epstopdf}           
\usepackage{latexsym}
\usepackage{times}

\newcommand{\cf}{\textit{cf.}~}
\newcommand{\ie}{\textit{i.e.,}~}
\newcommand{\eg}{\textit{e.g.,}~}

\newcommand{\hz}{{\rm \,Hz}}

\newcommand{\msun}{\,M_{\odot}}
\newcommand{\pls}{~~~}

\newcommand{\bea}{\begin{eqnarray}}
\newcommand{\eea}{\end{eqnarray}}
\newcommand{\beq}{\begin{equation}}
\newcommand{\eeq}{\end{equation}}

\begin{document}

\title{Vacuum Electromagnetic Counterparts of Binary Black-Hole Mergers}

\author{Philipp M\"osta}
\affiliation{
  Max-Planck-Institut f\"ur Gravitationsphysik,
  Albert-Einstein-Institut,
  Potsdam-Golm, Germany
}

\author{Carlos Palenzuela}
\affiliation{
  Canadian Institute of Theoretical Astrophysics,
  Toronto, Ontario, Canada
}
\affiliation{
  Max-Planck-Institut f\"ur Gravitationsphysik,
  Albert-Einstein-Institut,
  Potsdam-Golm, Germany
}

\author{Luciano Rezzolla}
\affiliation{
  Max-Planck-Institut f\"ur Gravitationsphysik,
  Albert-Einstein-Institut,
  Potsdam-Golm, Germany
}

\author{Luis Lehner}
\affiliation{
 Perimeter Institute for Theoretical Physics,
 Waterloo, Ontario, Canada
}
\affiliation{
 Department of Physics,
 University of Guelph,
 Guelph, Ontario, Canada
}
\affiliation{
Canadian Institute for Advanced Research, Cosmology \& Gravity Program
}

\author{Shin'ichirou Yoshida}
\affiliation{
Department of Earth Science and Astronomy,
Graduate School of Arts and Sciences,
University of Tokyo
}
\affiliation{
  Max-Planck-Institut f\"ur Gravitationsphysik,
  Albert-Einstein-Institut,
  Potsdam-Golm, Germany
}

\author{Denis Pollney}
\affiliation{
  Max-Planck-Institut f\"ur Gravitationsphysik,
  Albert-Einstein-Institut,
  Potsdam-Golm, Germany
}

\date{\today}

%
%
\begin{abstract}
As one step towards a systematic modeling of the electromagnetic (EM)
emission from an inspiralling black hole binary we consider a simple
scenario in which the binary moves in a uniform magnetic field
anchored to a distant circumbinary disc. We study this system by
solving the Einstein-Maxwell equations in which the EM fields are
chosen with strengths consistent with the values expected
astrophysically and treated as test-fields. Our initial data consists
of a series of binaries with spins aligned or anti-aligned with the
orbital angular momentum and we study the dependence of gravitational
and EM signals with different spin configurations. Overall we find
that the EM radiation in the lowest $\ell=2, m=2$ multipole accurately
reflects the gravitational one, with identical phase evolutions and
amplitudes that differ only by a scaling factor. This is no longer
true when considering higher $\ell$ modes, for which the amplitude
evolution of the scaled EM emission is slightly larger, while the
phase evolutions continue to agree. We also compute the efficiency of
the energy emission in EM waves and find that it scales quadratically
with the total spin and is given by $E^{\rm rad}_{_{\rm EM}}/M \simeq
10^{-15} \left({M}/{10^8\ M_{\odot}}\right)^2 \left({B}/{10^4\ {\rm
    G}}\right)^2$, hence $13$ orders of magnitude smaller than the
gravitational energy for realistic magnetic fields. Although large in
absolute terms, the corresponding luminosity is much smaller than the
accretion luminosity if the system is accreting at near the Eddington
rate. Most importantly, this EM emission is at frequencies of $\sim
10^{-4}(10^8 M_{\odot}/M)\,\hz$, which are well outside those
accessible to astronomical radio observations. As a result, it is
unlikely that the EM emission discussed here can be detected
\textit{directly} and simultaneously with the gravitational-wave
one. However, indirect processes, driven by changes in the EM fields
behavior could yield observable events. In particular we argue that if
the accretion rate of the circumbinary disc is small and sufficiently
stable over the timescale of the final inspiral, then the EM emission
may be observable \textit{indirectly} as it will alter the accretion
rate through the magnetic torques exerted by the distorted magnetic
field lines.
\end{abstract}

\maketitle

%
%
\section{Introduction} 

Gravitational-wave (GW) astronomy promises to revolutionize our
understanding of a number of astrophysical systems. Several
Earth-based detectors (LIGO, Virgo, GEO) are already operating at
their designed sensitivities and will be further upgraded in the
coming years.  Additionally, space-borne detectors are being
considered and might become a reality in the coming decade.  The ability to
harness the information carried by GWs will soon provide a completely
new way to observe the universe around us.  These detectors, along
with increasingly sensitive electromagnetic (EM) telescopes, will
provide insights likely to affect profoundly our understanding of
fundamental physics and the cosmos (see 
e.g.~\cite{Sathyaprakash:2009xs,Bloom:2009vx}
for a recent and detailed discussions of the astrophysics and cosmology
that will be possible with the detection of GWs)

Among the most promising sources of detectable GWs are systems
composed of binary black holes which, as they come together and merge,
radiate copious amounts of energy in the form of gravitational
radiation. When these black holes are supermassive, \ie with masses
$M\gtrsim 10^6\, M_{\odot}$, the cosmological and astrophysical
conditions leading to their formation will be such that prior to the
merger they will be surrounded by a gas or plasma that could also
radiate electromagnetically. Indeed, within the context of galaxy
mergers, such a scenario will typically arise as the central black
hole in each of the colliding galaxies sinks towards the gravitational
center, eventually forming a binary. In such a process, the binary
will generically find itself inside a circumbinary disc which might be
a catalyst of observable emissions as it interacts with the black
holes~\cite{2002ApJ...567L...9A,2003MNRAS.340..411L,Milosavljevic:2004cg}.
Within this context, several possibilities are actively being
investigated.  Among these, several studies have (with varying degrees
of approximations) concentrated on understanding emissions by the disc
due to the interaction with a recoiling, or a mass-reduced final black
hole~\cite{Schnittmann:2008,Lippai:2008,Shields:2008,Megevand2009,massloss,Corrales:2009nv}
and remnant gas around the black holes
~\cite{2008ApJ...672...83M,chang,vanMeter:2009gu,Bode:2009mt}.

A further intriguing possibility for such interaction is through EM
fields and in particular through the gravitomagnetic deformation of
magnetic fields, anchored to the disc, around the central region where
the black holes inspiral and merge. As the black holes proceed in an
ever shrinking orbit towards their ultimate coalescence, they will
twist and stir the EM fields and thus affect their topology.
Moreover, the spacetime dynamics might impact the fields in such a
strong way as to generate EM energy fluxes which may reach and impact
the disc and/or affect possible gas in the black holes' vicinity.
Last, but certainly not least, as the merger takes place the system
could acquire (at least at much later times) a configuration where
emissions through the Blandford-Znajek mechanism may take
place~\cite{Blandford1977}

The initial step towards understanding this system was taken
in~\cite{Palenzuela:2009yr,Palenzuela:2009hx} and highlighted the
possible phenomenology in the system. Although this first
investigation was restricted to the simplest case of an equal-mass,
nonspinning binary, it provided a proof of principle examination of
possible EM counterparts resulting from binary mergers. The work
presented here extends the study carried out
in~\cite{Palenzuela:2009yr,Palenzuela:2009hx} on several fronts.
First, it considers an broader class of binaries, with and without net
spins which are aligned/counter-aligned with the orbital angular
momenta (in the case of supermassive black holes, these configurations
are indeed the most likely to be
produced~\cite{Bogdanovic:2007hp,Dotti:2009vz}). Second, it studies
binaries from larger separations, thus allowing for a clearer modeling
of the inspiral. Third, it examines more closely the energy emissions
in both channels and compares the total output's dependence on final
black hole spin.  Finally, it assesses the direct detectability of the
EM radiation produced, and considers the astrophysical impact it may
have on the surrounding accretion disc. In addition, while sharing the
same astrophysical scenario, the methodological approach adopted here
also differs from that of~\cite{Palenzuela:2009yr,Palenzuela:2009hx}
in at least three important ways. First, we here explicitly impose the
test-field limit by setting to zero the stress-energy tensor on the
right-hand-side of the Einstein equations. Second, we employ a
different formulation of Einstein equations, \ie a conformal
transverse-traceless one in place of a generalized-harmonic
one. Third, we use a distinct computational infrastructure based on
\texttt{Cactus} and \texttt{Carpet} rather than the \texttt{HAD} code
which was instead used
in~\cite{Palenzuela:2009yr,Palenzuela:2009hx}). The excellent
agreement in the phenomenology observed, in cases considered in both
works, gives a strong further indication of the correctness of the
results obtained.

Our analysis shows that the EM-wave emission in the lowest $\ell=2,
m=2$ multipole reflects in a spectacular way the gravitational one and
that the phase and amplitude evolution differ only by a scaling
factor. This is no longer true when considering higher $\ell$ modes,
for which the amplitude evolution of the EM emission is slightly
larger. We also find that the efficiency of the energy emission in EM
waves scales quadratically with the total spin and is given, for
realistic magnetic fields, by $E^{\rm rad}_{_{\rm EM}}/M \simeq
10^{-15} \left({M}/{10^8\ M_{\odot}}\right)^2 \left({B}/{10^4\ {\rm
    G}}\right)^2$. For expected fields of $10^4$ G this energy loss is
$13$ orders of magnitude smaller than the gravitational one. In
addition, the corresponding EM luminosity is much smaller than that of
accretion, if the system is accreting at near the Eddington rate. Most
importantly, this emission is at frequencies of $\sim 10^{-4}(10^8
M_{\odot}/M)\,\hz$, which lies outside the range of astronomical radio
observations. As a result, it is highly unlikely that the EM emission
discussed here can be detected directly and simultaneously with the GW
one. Other processes however could be affected by this flux of EM
energy and produce detectable effects. For instance, if the accretion
rate of the circumbinary disc is small and sufficiently stable over
the timescale of the final inspiral, then the EM emission may be, in
particular, observable indirectly as it will alter the accretion rate
through the magnetic torques exerted by the distorted magnetic field
lines.

This work is organized as follows: Sect.~\ref{sec:evolution_eqs}
presents an overview of both Maxwell and Einstein equations as
implemented within the BSSNOK formulation of the Einstein
equations. The physical and astrophysical setup adopted in our
simulations is discussed in Sect.~\ref{sec:physical_setup}, while
Sect.~\ref{sec:single_BHs} is dedicated to the analysis of the EM
fields for isolated black holes, either nonspinning or with spin
aligned with the orbital angular momentum. Sect.~\ref{sec:BBHs}
collects instead our results for the different binaries considered,
while the assessment of the astrophysical impact of the EM emission is
presented in Sect.~\ref{sec:astrophysics}. We conclude in
Sect.~\ref{sec:conclusions} with final comments and discussions.

%
%
%
\section{The Evolution Equations} 
\label{sec:evolution_eqs}

We solve the Einstein-Maxwell system to model the interaction of an
inspiralling black-hole binary with an externally sourced magnetic
field in an electro-vacuum spacetime. More specifically, we solve the
Einstein equations 
\begin{eqnarray}
  R_{\mu\nu} - \frac{1}{2} {} Rg_{\mu\nu}&=& 8\pi T_{\mu\nu}  \label{E1} ~~,~~\
\end{eqnarray}
where $g_{\mu \nu}$, $R_{\mu \nu}$ and $T_{\mu\nu}$ are the metric,
the Ricci and the stress-energy tensor,
respectively, together with an extension of Maxwell
equations in absence of currents~\cite{2007MNRAS.382..995K,Palenzuela2008},
\begin{eqnarray}
  \nabla_{\mu} (F^{\mu \nu} + g^{\mu \nu} \Psi) &=& - \kappa\, t^{\nu} \Psi \, ,
  \label{Maxwell1a} \\
  \nabla_{\mu} (^*F^{\mu \nu} + g^{\mu \nu} \phi) &=& -\kappa\, t^{\nu} \phi \, ;
\label{Maxwell1b}
\end{eqnarray}
when written as conservation laws for the Faraday tensor $F_{\mu \nu}$
and of its dual ${}^*F_{\mu \nu}$. Note that we have here introduced
two extra scalar fields $\Psi$ and $\phi$, which are initially zero
and whose evolution drives the system to a satisfaction of the EM
constraints (see discussion below). The two systems are
coupled through the stress-energy tensor,
\begin{eqnarray}\label{stress-em}
   T_{\mu \nu} = \frac{1}{4\, \pi} \left[ {F_{\mu}}^{\lambda} ~ F_{\nu
       \lambda} - \frac{1}{2}\, g_{\mu \nu} ~ F^{\lambda \sigma}
     F_{\lambda \sigma} \right] \, .
\end{eqnarray}
Note also that since our stress-energy tensor is not identically zero,
the binary is not in vacuum, at least within a strict
general-relativistic sense. However, because the sources of the EM
fields are not part of our description and we simply consider EM
fields, our setup will be that of an ``electromagnetic-vacuum'', which
will simply refer to as ``vacuum''.

\subsection{The Einstein Equations}

The numerical solution of the Einstein equations has been performed
using a three-dimensional finite-differencing code solving a
conformal-traceless ``$3+1$'' BSSNOK formulation of the Einstein
equations (see~\cite{Pollney:2009yz} for the full expressions in
vacuum and~\cite{Baiotti08} for spacetimes with matter) using the
\texttt{Cactus} Computational Toolkit~\citep{cactusweb} and the
\texttt{Carpet}~\citep{Schnetter-etal-03b} adaptive mesh-refinement
driver. Recent developments, such as the use of $8$th-order
finite-difference operators or the adoption of a multiblock
structure to extend the size of the wave zone have been recently
presented in~\cite{Pollney:2009ut,Pollney:2009yz}. Here, however, to
limit the computational costs and because a very high accuracy in the
waveforms is not needed, the multiblock structure was not used. For
compactness we will not report here the details of the formulation of
the Einstein equations solved or the form of the gauge conditions
adopted. All of these aspects are discussed in detail
in~\cite{Pollney:2009yz}, to which we refer the
interested reader.

\subsection{The Maxwell Equations}

Maxwell equations~\eqref{Maxwell1a}--\eqref{Maxwell1b} take a more
familiar form when represented in terms of the standard electric and
magnetic fields. These are defined by the following decomposition of
the Faraday tensor
\begin{eqnarray}\label{Faraday tensor}
  F^{\mu \nu} &=& t^{\mu} E^{\nu} - t^{\nu} E^{\mu}
               + \epsilon^{\mu\nu\alpha\beta}~B_{\alpha}\,t_{\beta}
\label{F_em1a} \, , \\
  ^*F^{\mu \nu} &=& t^{\mu} B^{\nu} - t^{\nu} B^{\mu}
               - \epsilon^{\mu\nu\alpha\beta}~E_{\alpha}\,t_{\beta} \, ;
\label{F_em1b}
\end{eqnarray}
where $t^{\mu}$ is the unit time vector associated with a generic normal
observer to the hypersurfaces. The vectors $E^{\mu}$ and $B^{\mu}$ are
the (purely spatial, $E^{\mu} t_{\mu} = B^{\mu} t_{\mu} = 0$)
electric and magnetic fields measured by such observer.

As mentioned above, we adopt an extended version of Maxwell's equations 
which introduces two extra scalar fields $\Psi$ and $\Phi$. This extension
induces evolution equations for the EM constraints ($\nabla_i E^i = 0
=\nabla_i B^i$) described by damped wave equations and so control
dynamically these constraints. In terms of $E^{\mu}$ and $B^{\mu}$ the
$3+1$ version of (\ref{Maxwell1a}-\ref{Maxwell1b}) results,
\begin{eqnarray}
&& \hskip -0.5 cm {\cal D}_t \, E^{i}  -
  \epsilon^{ijk} \nabla_j (\,\alpha\ B_c\,)
   + \alpha\, \gamma^{ij} \nabla_j\,\Psi = \alpha\, K\, E^i\,,  \\
\label{maxwellext_$3+1$_eq1a} 
&& \hskip -0.5 cm {\cal D}_t \, B^{i} +
  \epsilon^{ijk} \nabla_j (\,\alpha\, E_c\,) 
  + \alpha\, \gamma^{ij} \nabla_j\, \Phi = \alpha\, K\, B^i\,, \\
\label{maxwellext_$3+1$_eq1b} 
&& \hskip -0.5 cm {\cal D}_t \,\Psi + \alpha\, \nabla_i E^i =
   -\alpha \kappa\, \Psi\,, \\
\label{maxwellext_$3+1$_eq1c} 
&& \hskip -0.5 cm {\cal D}_t \,\Phi + \alpha\, \nabla_i B^i =
   -\alpha \kappa\, \Phi \,.
\label{maxwellext_$3+1$_eq1d}
\end{eqnarray}
where ${\cal D}_t\equiv (\partial_t - {\cal L}_{\beta})$ and ${\cal
  L}_{\beta}$ is the Lie derivative along the shift vector
$\boldsymbol{\beta}$. Exploiting that the covariant derivative
in the second term of (\ref{maxwellext_$3+1$_eq1a} - \ref{maxwellext_$3+1$_eq1b}) reduces to a partial one
\begin{equation}
\epsilon^{ijk} \nabla_j B_k = \epsilon^{ijk} (\partial_j + \Gamma^c_{kj} B_k) = \epsilon^{ijk} \partial_j B_k, 
\label{partial_eq1}
\end{equation}
and using a standard conformal decomposition of the spatial 3-metric 
\begin{eqnarray}
\tilde\gamma_{ij} = e^{4\phi} \gamma_{ij} \,\, \, ; \,\,\,
\label{conformal_eq1}
\phi = \frac{1}{12}\mathrm{ln}\gamma
\label{conformal_eq2}
\end{eqnarray}
one obtains the final expressions
\begin{widetext}
\begin{eqnarray}
&& \hskip -0.5 cm {\cal D}_t \, E^{i}  -
   \epsilon^{ijk}\, e^{4\phi}\, [\,(\partial_j\, \alpha\,)\, \tilde\gamma_{ck}\, B^c\,  
   + \alpha\,(\,4\,\tilde\gamma_{ck}\,\,\partial_j\,\phi\,+\,\partial_j\,\tilde\gamma_{ck}\,)\,B^c 
+ \alpha\,\tilde\gamma_{ck}\,\partial_j\,B^c\, ] + 
   \alpha\,e^{-4\phi}\,\tilde\gamma^{ij}\,\nabla_j\,\Psi = \alpha\, K\, E^i\,,  \\
\label{maxwellext_$3+1$_p_eq1a} 
&& \hskip -0.5 cm {\cal D}_t \, B^{i}  +
   \epsilon^{ijk}\, e^{4\phi}\, [\,(\partial_j\, \alpha\,)\, \tilde\gamma_{ck}\, E^c\,  
   + \alpha\,(\,4\,\tilde\gamma_{ck}\,\,\partial_j\,\phi\,+\,\partial_j\,\tilde\gamma_{ck}\,)\,E^c 
+ \alpha\,\tilde\gamma_{ck}\,\partial_j\,E^c\, ] + 
   \alpha\,e^{-4\phi}\,\tilde\gamma^{ij}\,\nabla_j\,\Phi = \alpha\, K\, B^i\,,  \\
\label{maxwellext_$3+1$_p_eq1b} 
&& \hskip -0.5 cm {\cal D}_t \,\Psi + \alpha\, \nabla_i E^i =
   -\alpha \kappa\, \Psi\,, \\
\label{maxwellext_$3+1$_p_eq1c} 
&& \hskip -0.5 cm {\cal D}_t \,\Phi + \alpha\, \nabla_i B^i =
   -\alpha \kappa\, \Phi \,.
\label{maxwellext_$3+1$_p_eq1d}
\end{eqnarray}
\end{widetext}
Notice that the standard Maxwell equations in a curved background
are recovered for $\Psi=\Phi=0$. The $\Psi$ and $\Phi$ scalars can
then be considered as the normal-time integrals  of the standard
divergence constraints
\begin{equation}\label{Maxwell_divs}
    \nabla_i E^i =  0 \,\, , \, \, \nabla_i B^i = 0
\end{equation}
These constraints propagate with light speed and
are damped during the evolution.

As mentioned above, the coupling between the Einstein and Maxwell
equations takes place via the inclusion of a nonzero stress-energy
tensor (\cf the set of equations presented in~\cite{Baiotti08}) for
the EM fields and which is built in terms of the Faraday tensor as
described in~\eqref{stress-em}. More specifically, the relevant
components of the stress-energy tensor can be obtained in terms of the
electric and magnetic fields, that is
\begin{eqnarray}\label{Tmunu_decomposition2}
   \tau &=& \frac{1}{8\pi} (E^2 + B^2) ~~~,~~~
   S_{i} = \frac{1}{4 \pi}\epsilon_{ijk} E^j B^k ~~~,~~~ \\
   S_{ij} &=& \frac{1}{4 \pi}\left[-E_i E_j - B_i B_j + \frac{1}{2}\,
    \gamma_{ij}\, (E^2 + B^2)\right]
\end{eqnarray}
where $E^2 \equiv E^k E_k$ and $B^2 \equiv B^k B_k$. The scalar
component $\tau$ can be identified with the energy density of the EM
field (\ie $\rho_{_{\rm ADM}}$ in~\cite{Baiotti08}) and the energy
flux $S_i$ is the Poynting vector. As already discussed in the
Introduction, we stress again that the EM energies considered here are
so small when compared with the gravitational ones that the
contributions of the stress-energy tensor to the right-hand-side of
the Einstein equations~\eqref{E1} are comparatively negligible and
thus effectively set to zero\footnote{The fully coupled set of the
  Einstein-Maxwell equations was considered
  in~\cite{Palenzuela:2009yr,Palenzuela:2009hx} and the comparison
  with the results obtained here suggests that for the fields below
  $10^8$ G, the use of the test-field approximation is justified.}

\subsection{Analysis of Radiated Quantities}

The calculation of the EM and gravitational radiation generated during
the inspiral, merger and ringdown is arguably the most important
aspect of this work as it allows us to establish on a firm basis
whether a clear correlation should be expected between the two forms
of radiation. We compute the gravitational radiation via the
Newman-Penrose curvature scalar $\Psi_4$ defined as
\begin{equation}
  \Psi_4 \equiv -C_{\alpha\beta\gamma\delta}
    n^\alpha \bar{m}^\beta n^\gamma \bar{m}^\delta,
  \label{eq:psi4def}
\end{equation}
where $C_{\alpha\beta\gamma\delta}$ is the Weyl curvature tensor,
which is projected onto a null frame, $\{\boldsymbol{l},
\boldsymbol{n}, \boldsymbol{m}, \bar{\boldsymbol{m}}\}$. In practice,
we define an orthonormal basis in the three space
$(\hat{\boldsymbol{r}}, \hat{\boldsymbol{\theta}},
\hat{\boldsymbol{\phi}})$, centered on the Cartesian grid center and
oriented with poles along $\hat{\boldsymbol{z}}$. The normal to the
slice defines a time-like vector $\hat{\boldsymbol{t}}$, from which we
construct the null frame
\begin{equation}
   \boldsymbol{l} = \frac{1}{\sqrt{2}}(\hat{\boldsymbol{t}} - \hat{\boldsymbol{r}}),\quad
   \boldsymbol{n} = \frac{1}{\sqrt{2}}(\hat{\boldsymbol{t}} + \hat{\boldsymbol{r}}),\quad
   \boldsymbol{m} = \frac{1}{\sqrt{2}}(\hat{\boldsymbol{\theta}} - 
     {\mathrm i}\hat{\boldsymbol{\phi}}) \ .
\end{equation}
We then calculate $\Psi_4$ via a reformulation of (\ref{eq:psi4def}) 
in terms of ADM variables on the slice~\cite{Shinkai94, Pollney:2009yz},
\begin{equation}
  \Psi_4 = C_{ij} \bar{m}^i \bar{m}^j,  \label{eq:psi4_adm}
\end{equation}
where 
\begin{equation}
  C_{ij} \equiv R_{ij} - K K_{ij} + K_i{}^k K_{kj} 
    - {\rm i}\epsilon_i{}^{kl} \nabla_l K_{jk}.
\end{equation}
We note that the we have also implemented an independent method to
compute gravitational radiation via the measurements of the
nonspherical gauge-invariant metric perturbations of a Schwarzschild
black hole~\cite{Abrahams97a_shortlist,Rupright98,Rezzolla99a} (A
review on the basic formalism can be found in ~\cite{Nagar05} and
refs.~\cite{Abrahams97a_shortlist,Rupright98,Rezzolla99a} provide
examples of the applications of this method to Cartesian-coordinate
grids). For compactness, hereafter we will limit our discussion to the
gravitational radiation measured in terms of the curvature scalar
$\Psi_4$.

In a similar manner, the EM radiation can be measured using the
Newman-Penrose scalar $\Phi_2$ defined as
\begin{equation}
\Phi_2 \equiv F^{\mu\nu} \bar m_{\mu} n_{\nu}\, , \label{radiation}
\end{equation}
with the same same tetrad used for $\Psi_4$, allowing to measure
outgoing EM radiation. (Possible gauge effects, as those discussed
in~\cite{lehnermoreschi}, have been seen to be negligible
in~\cite{Palenzuela:2009hx}, and the results here agree with those).

Using the curvature scalars $\Psi_4$ and $\Phi_2$ it is also possible
to compute the energy carried off by outgoing waves at infinity. More
specifically, the total energy flux per unit solid angle can be
computed directly as
\begin{eqnarray}
  F_{_{\rm GW}} &=& \frac{{dE}_{_{\rm GW}}}{dt\, d\Sigma} = 
\lim_{r \rightarrow \infty} \frac{r^2}{16 \pi} \left| \int_{- \infty}^t \Psi_4 dt' \right|^2
\label{FGW} \\
  F_{_{\rm EM}} &=& \frac{{dE}_{_{\rm EM}}}{dt\, d\Sigma} = \lim_{r \rightarrow \infty} 
                                       \frac{r^2}{4 \pi} |\Phi_2|^2 ~~.
\label{FEM}
\end{eqnarray}
Representing now $\Psi_4$ and $\Phi_2$ via a standard decomposition
into spherical harmonics using the spin-weights $-2$ for $\Psi_4$ and
$-1$ for $\Phi_2$, we obtain the final expressions
\begin{eqnarray}
  F_{_{\rm GW}} &=& \frac{{dE}_{_{\rm GW}}}{dt} = 
\lim_{r \rightarrow \infty} \frac{r^2}{16 \pi} \sum_{l,m} \left|\int_{-\infty}^t A^{lm}\right|^2 dt' 
\label{FGW_modes} \\
  F_{_{\rm EM}} &=& \frac{{dE}_{_{\rm EM}}}{dt} = \lim_{r \rightarrow \infty} 
                                       \frac{r^2}{4 \pi} \sum_{l,m} \left|B^{lm}\right|^2 ~~,
\label{FEM_modes}
\end{eqnarray}
where $A^{lm}$ and $B^{lm}$ are the coefficients of the spherical 
harmonic decomposition of $\Psi_4$ and $\Phi_2$, respectively.

%
%
\section{Physical and Astrophysical Setup} 
\label{sec:physical_setup}

As mentioned in the introduction, the astrophysical scenario we have
in mind is motivated by the merger of supermassive black holes
binaries resulting from galaxy mergers. More specifically, we consider
the astrophysical conditions that would follow the merger of two
supermassive black holes, each of which is surrounded by an accretion
disc. As the merger between the two galaxies takes place and the black
holes become close, a ``circumbinary'' accretion forms and reaches a
stationary accretion phase. During this phase, the binary evolves on
the timescale of the emission of gravitational radiation and its
separation progressively decreases as gravitational waves carry away
energy and angular momentum from the system. This
radiation-reaction timescale is much longer than the (disc) accretion
timescale, which is regulated by the ability of the disc to
transport outwards its angular momentum (either via viscous shear or
magnetically-mediated instabilities). As a consequence, for most of the evolution 
the disc slowly follows the binary as its orbit shrinks. However, as the
binary separation becomes of the order of $\sim 10^5-10^6\,M$, the
radiation-reaction timescale reduces considerably and can become
smaller than the disc accretion one. When this happens, the disc
becomes disconnected from the binary, the mass accretion rate reduces
substantially and the binary performs its final orbits in an
``interior'' region which is essentially devoid of
gas~\cite{2002ApJ...567L...9A,2003MNRAS.340..411L,Milosavljevic:2004cg}. This
represents the astrophysical setup of our simple model.

We introduce a coupling between the binary and the disc via a
large-scale magnetic field which we assume to be anchored to the disc,
whose inner edge is at a distance of $\sim 10^3\, M$ and is
effectively outside of our computational domain, while the binary
separation is only of $\sim 10\,M$, where $M$ is the total
gravitational mass of the binary. We note that although the
large-scale magnetic field is poloidal, it will appear as essentially
uniform within the ``interior region'' where the binary evolves and
which we model here. As a result, the initially magnetic field adopted
has Cartesian components given simply by $B^i = (0,0,B_0)$ with
$B_0\,M = 10^{-4}$ in geometric units or $B_0\sim 10^8$ G for a binary
with total mass $M=10^8\,M_{\odot}$. Furthermore, because we consider
an electromagnetic vacuum, the charges, electric currents and the
initial electric field are all assumed to be zero, \ie $E^i = 0$.

We note that although astrophysically large, the initial magnetic
field considered here has an associated EM energy which is several
orders of magnitude smaller than the gravitational-field energy. As a
result, any effect from the EM field dynamics on the spacetime itself
will be negligible and so the EM fields are treated here as
test-fields.  The case of stronger magnetic fields and their
consequent impact on the spacetime will be presented in a forthcoming
work.

%
%
\section{Isolated Black Holes}
\label{sec:single_BHs}

We first study isolated black holes, both as a check of our
implementation and to analyze the interaction of the chosen external
initial magnetic field with the spacetime curvature generated by the
black holes. The initial magnetic field in all simulations is uniform
with strength $B_0$ and aligned with the $z$-axis, while the initial
electric field is zero everywhere. Although this solution satisfies
the Maxwell equations trivially, it is not a stationary solution of the coupled
Einstein-Maxwell system for the chosen black hole initial data. The
solution thus exhibits a transient behavior and evolves towards a
time-independent state given by a solution first found by
Wald~\cite{Wald:74bh}. One important feature of Wald's solution is
that in the case of spinning black holes, a net charge (and hence a
net electric field) will develop as a result of ``selective
accretion'' and whose asymptotic value is simply given by
$Q=2B_0J$. Although this charge is astrophysically uninteresting,
being limited to be $Q/M \leq 2B_0M \simeq 1.7\times 10^{-20}\, B_0
(M/M_{\odot})$ G~\cite{Wald:74bh} for a Kerr black hole with
$J/M^2\leq 1$, it represents an excellent testbed for our numerical
setup.

\begin{figure}[t]
\begin{center}
   \scalebox{0.4}{\includegraphics[angle=-0]{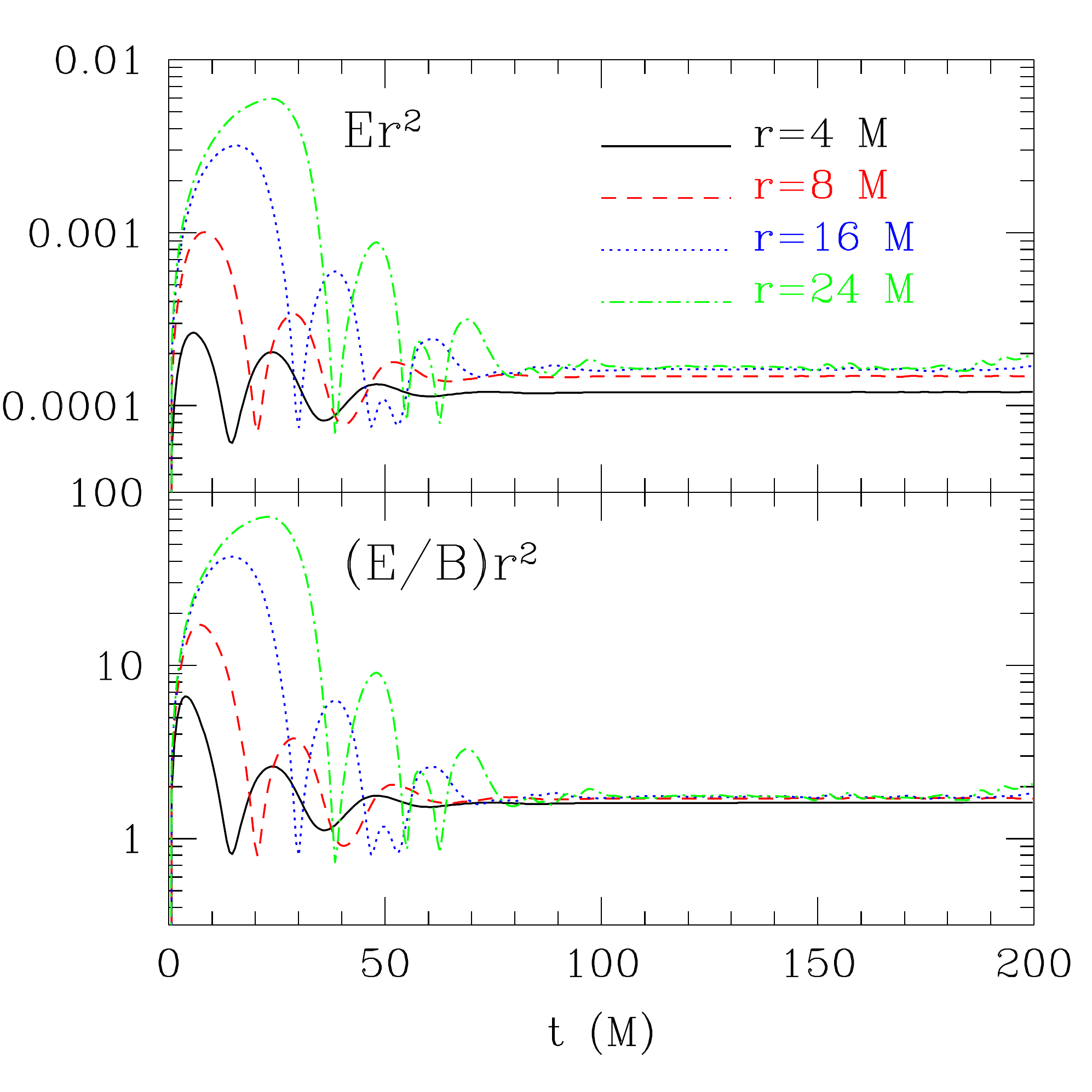}}
  \vskip -0.5cm
  \caption{Recovery of Wald's solution for an isolated Kerr black hole
    with dimensionless spin $a=0.7$. Shown in the top panel are the
    values of the electric field as measured at different distances
    from the origin; since $E r^2 \sim B_0 J$, the different lines
    should overlap at late times if the magnetic field is uniform
    which is evident in the figure. Shown in the bottom panel is the
    ratio of the electric and magnetic fields which is
    proportional to the black hole spin only.  Note the transitory
    state until $t \approx 70M$, when the solution reaches a
    stationary state.}
  \label{fig:tau}
\end{center}
\end{figure}

To validate the ability of the code to recover this analytic solution
we have performed several tests involving either a Schwarzschild black
hole or Kerr black holes with dimensionless spin parameters
$a=J/M^2=0.7$ (this value chosen as it is close to the final spin values
resulting from the merger simulations covered in
Sect.~\ref{sec:single_BHs}). In this latter case, the spin vector
was chosen to be either parallel to the background magnetic field, \ie
with $J^i=(0, 0, J)$ or orthogonal to it, \ie with
$J^i=(J, 0, 0)$. As expected, the early stages in the
evolution reveal a transient behavior as the EM fields rearrange
themselves and adapt to the curved spacetime reaching a stationary
configuration after about $\sim 70\,M$. The electric field, in
particular, goes from being initially zero to being nonzero and
decaying radially from the black hole.

\begin{figure*}[t]
\begin{center}
   \hskip -0.5cm
   \scalebox{0.42}{\includegraphics[angle=-0]{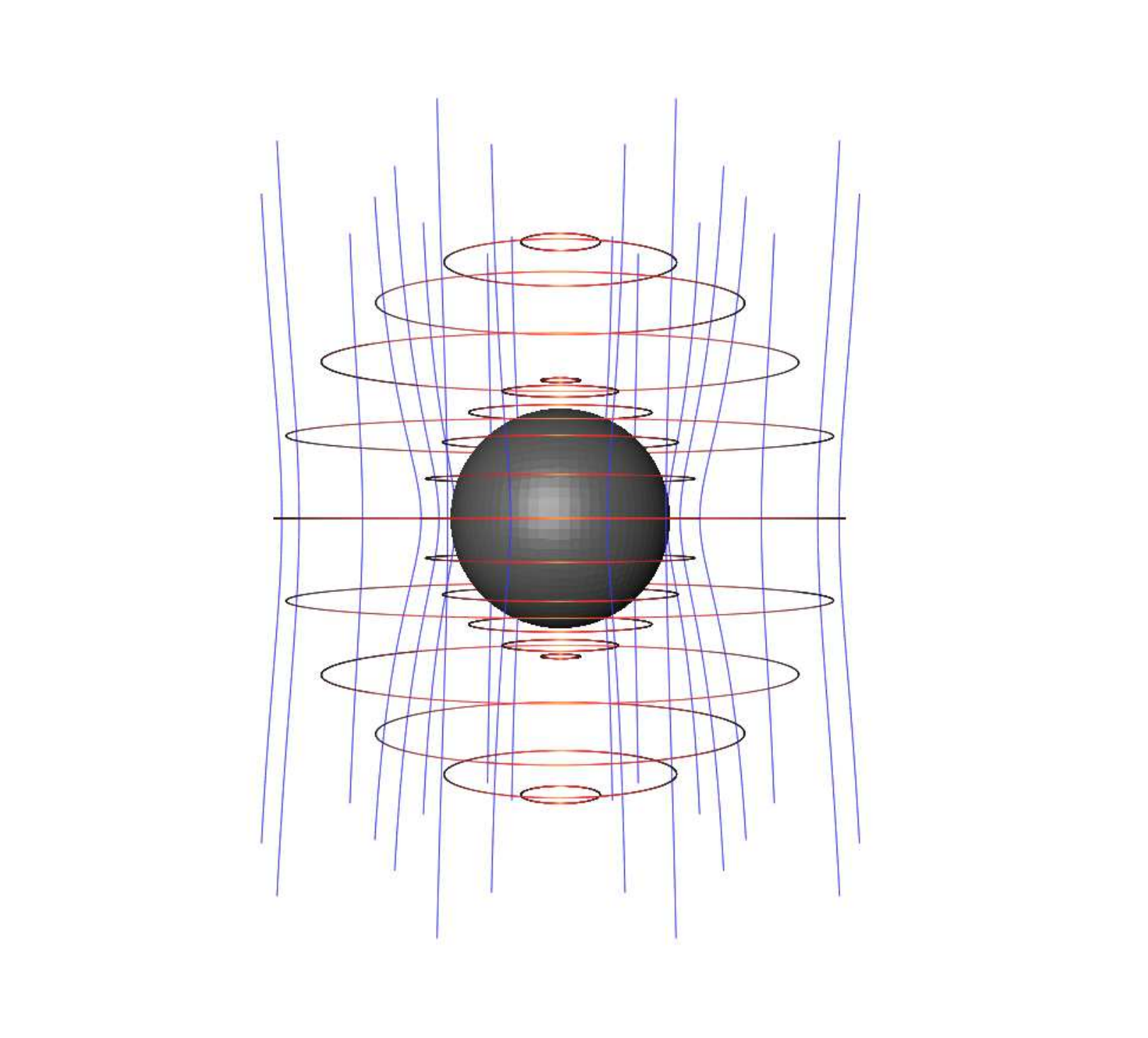}}
\hskip -1.cm
   \scalebox{0.42}{\includegraphics[angle=-0]{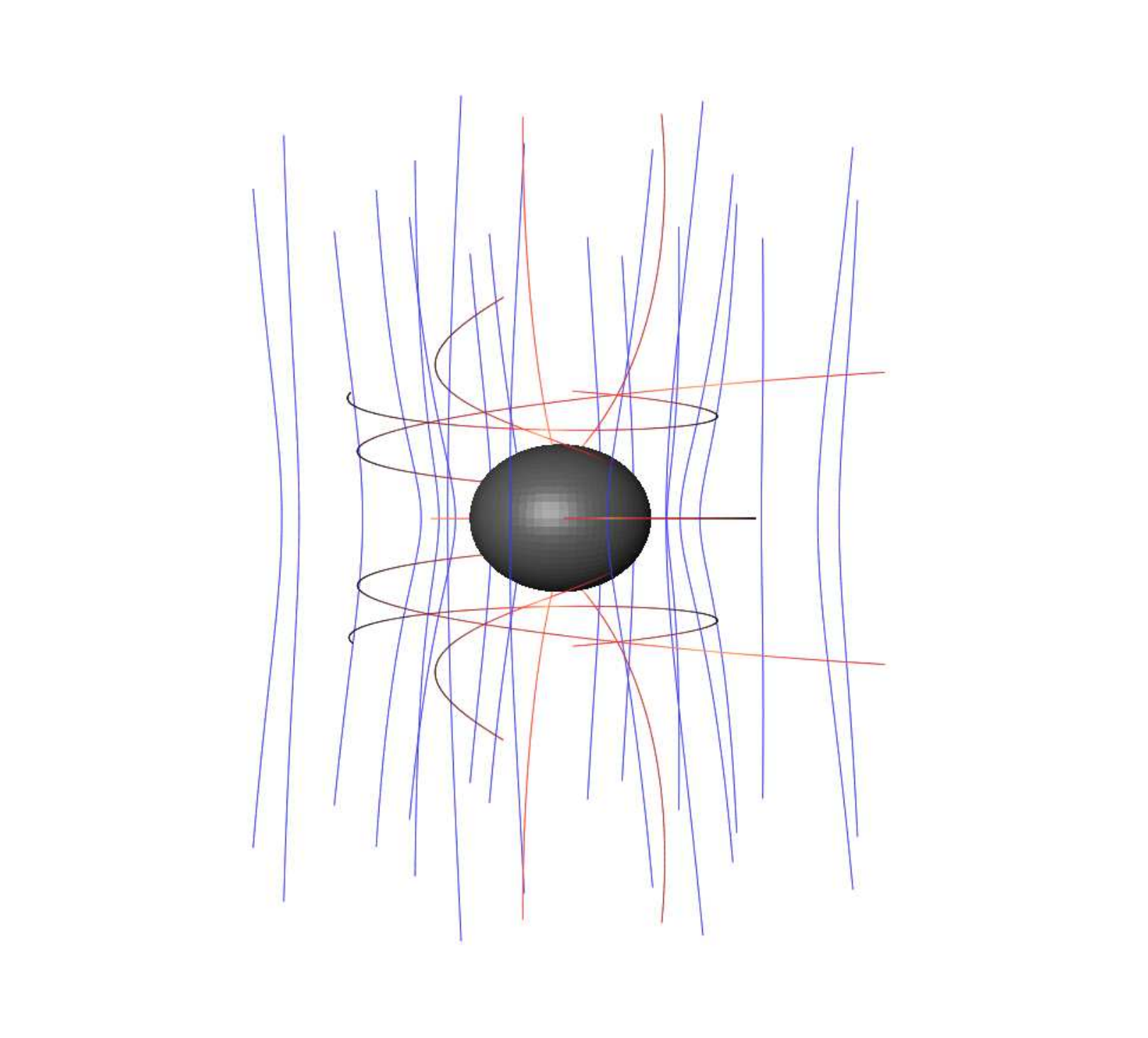}}
   \scalebox{0.42}{\includegraphics[angle=-0]{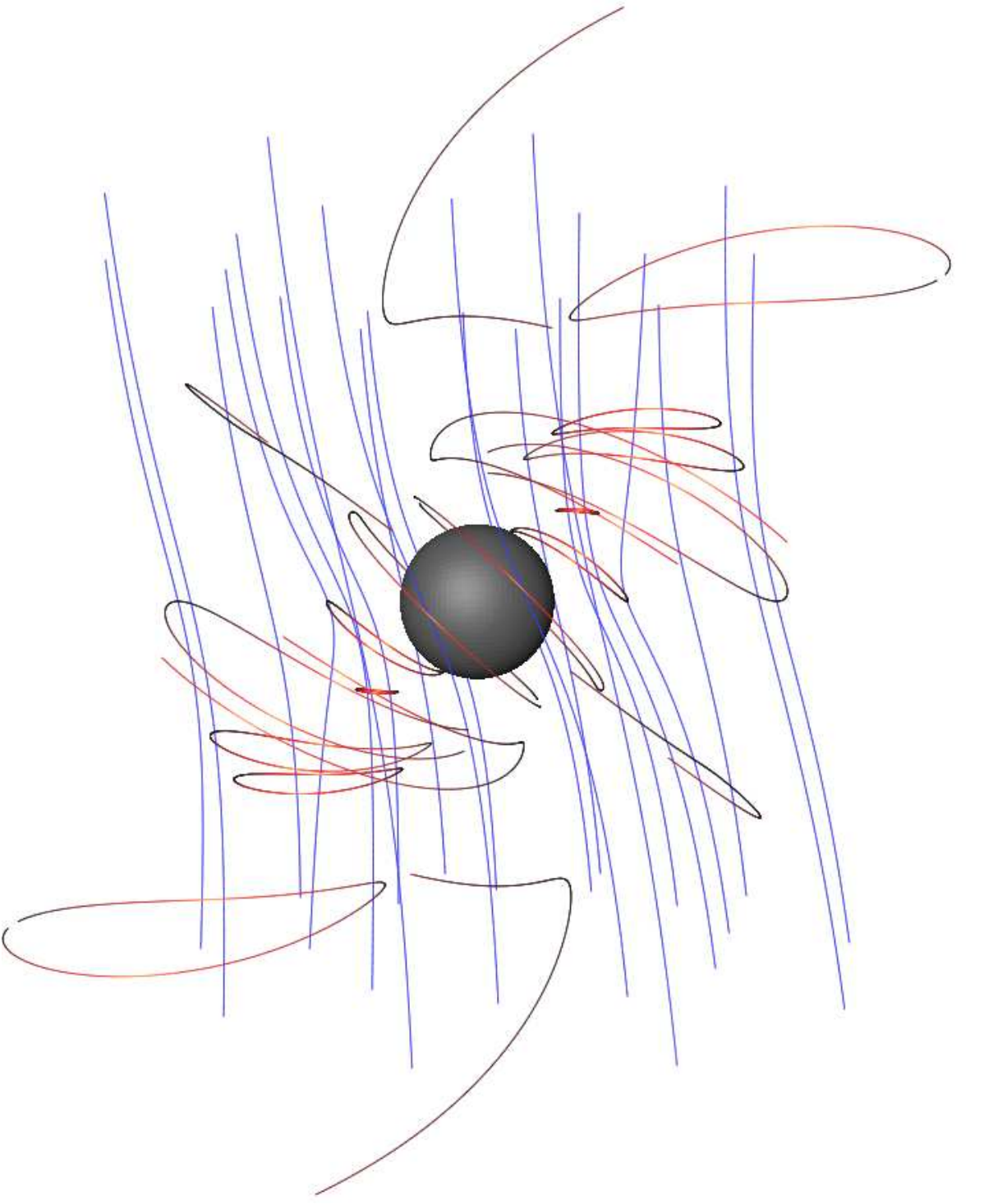}}
  \caption{\label{fig:fields_a0} \textit{Left panel:} Magnetic (blue)
    and electric field (red/magenta) field lines at $t=200\,M$ for a
    Schwarzschild black hole. \textit{Central panel:} the same as in
    the left panel but for a Kerr black hole with spin $a=J/M^2=0.7$
    aligned with the magnetic field, \ie $J^i=\left\{0, 0,
    J\right\}$. \textit{Right panel:} the same as in the center panel
    but for a Kerr BH with spin $a=J/M^2=0.7$ which is orthogonal to
    the magnetic field, \ie $J^i=\left\{J, 0, 0 \right\}$. Indicated
    with black surfaces are the apparent horizons.}
\end{center}
\end{figure*}

Although the original solution found by Wald was expressed in
Boyer-Lindquist coordinates, there is a simple way to validate that
our gauge is sufficiently similar (at least at far distances) and that
the numerical solution approaches Wald's one for an isolated black
hole in a uniform magnetic field. This is shown in Fig.~\ref{fig:tau},
which reports the time evolution of the EM fields $E$ and $B$ for a
simulation of the Kerr black hole with spin $a=0.7M$ aligned with the
magnetic field. In particular, the top panel shows the time evolution
of the electric field when the latter is rescaled by the radial
positions where it is measured, \ie $E r^2$ with $r=4\,M, 8\,M, 16\,M$
and $24\,M$. Because of $E \propto B_o J/r^2$, one expects the
different lines to be on top of each other.  This is clearly the case
for the data extracted at $r=16\,M$ and $24\,M$, but it ceases to be
true for the data at $r= 4\,M, 8\,M$, for which the magnetic field and
gauge structure are strongly influenced by the black-hole geometry.
Interestingly, however, in this strong-field region near the black
hole another scaling can be found and which is closely related to one
expressed by Wald's solution. In particular, the radial dependence of
the magnetic field can be factored out by considering the ratio of the
electric and magnetic field which, in Wald's solution, should be
proportional to the black-hole spin only. The bottom panel of
Fig.~\ref{fig:tau} shows therefore the evolution of $(E/B)r^2 \sim J$
which is indeed a constant at all the radial positions as shown by
the good overlap among the different curves.  We find that this
scaling can be used as an effective test which is valid at all radial
positions. These observations, together with the clear approach to a
stationary configuration indicate the asymptotic (in time)
behavior is indeed described by Wald's solution.

\begin{figure*}[t]
\begin{center}
   \scalebox{0.425}{\includegraphics[angle=-0]{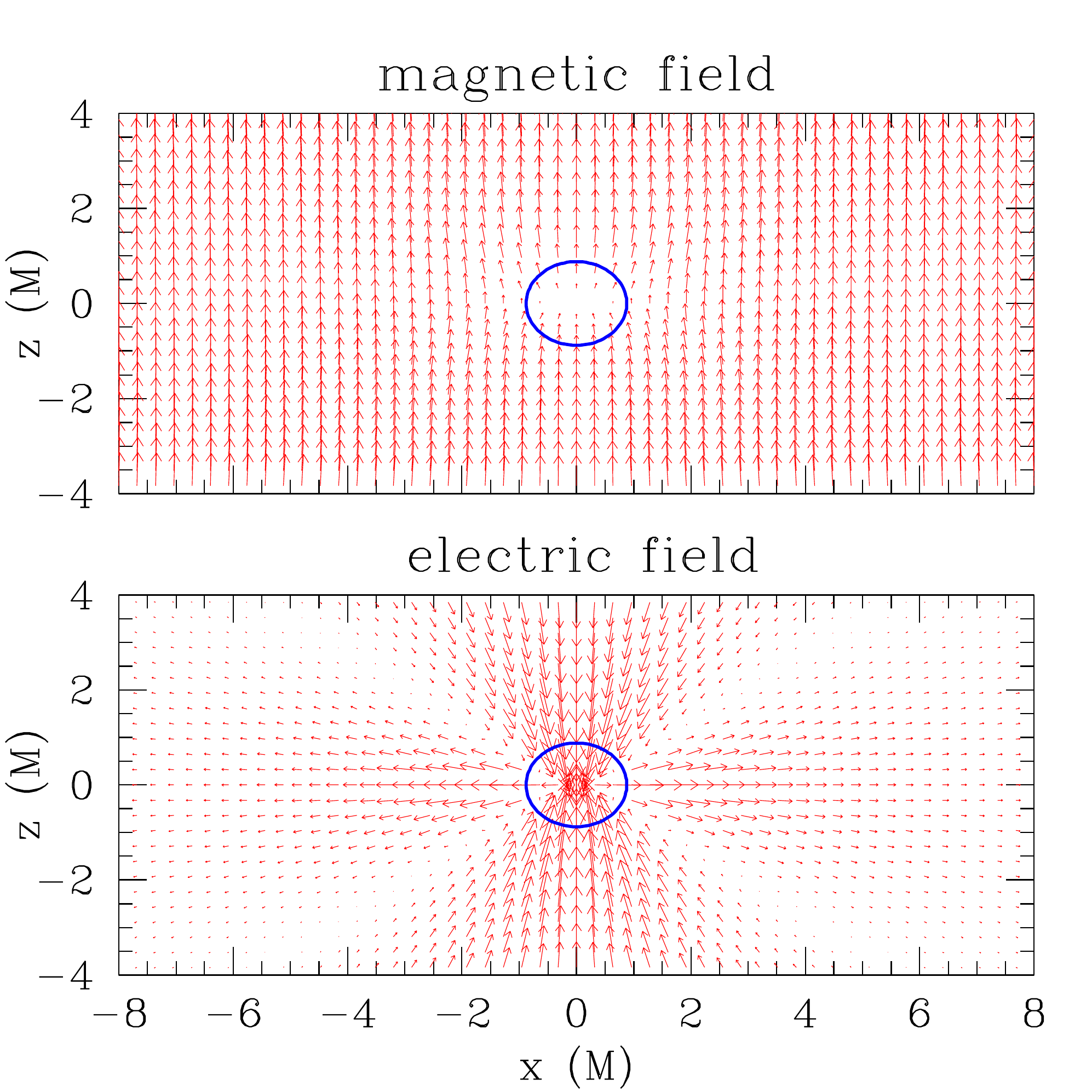}}
   \scalebox{0.425}{\includegraphics[angle=-0]{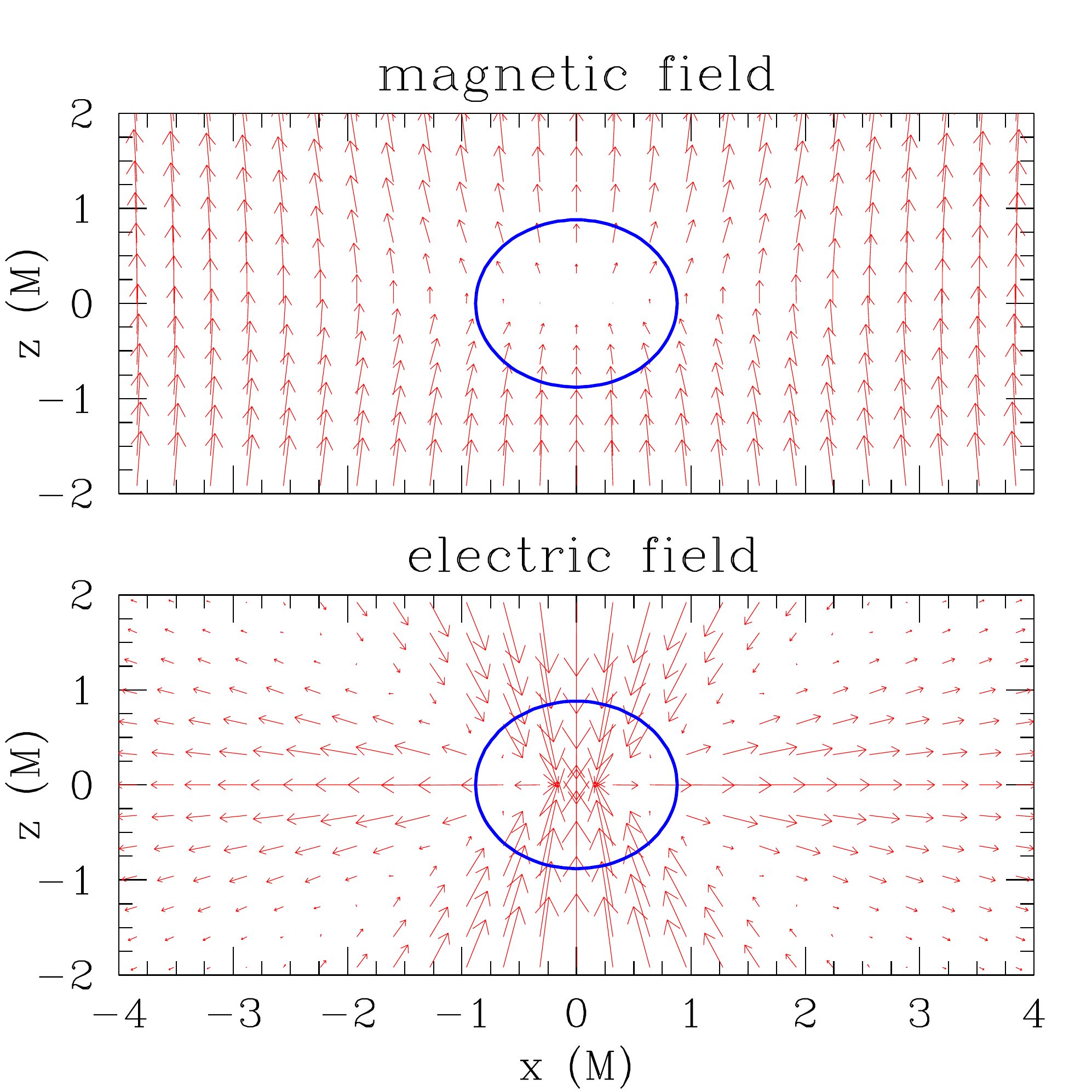}}
  \vskip -0.25cm
  \caption{\textit{Left panel:} Large-scale magnetic and electric
    field lines on the plane $y=0$ and at $t=200\,M$ for a Kerr black
    hole with spin $J/M^2=0.7$ aligned with the magnetic field, \ie
    along the $z$-axis. Indicated with blue circles are the apparent
    horizons. \textit{Right panel:} The same as on the left panel but
    on a smaller scale to highlight the fields structure in the
    vicinity of the black hole.}
  \label{fig:fields_a07_y0}
\end{center}
\begin{center}
   \scalebox{0.425}{\includegraphics[angle=-0]{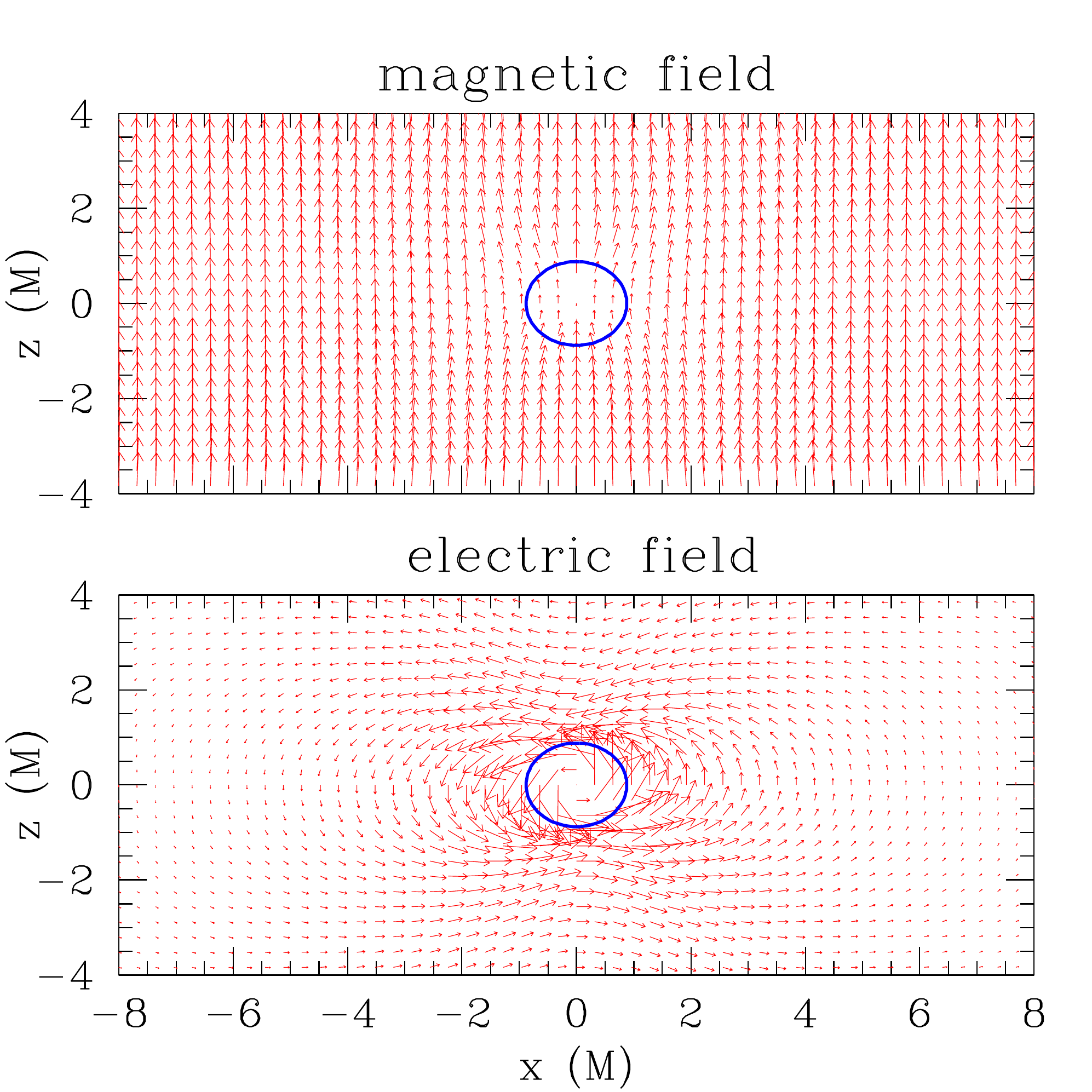}}
   \scalebox{0.425}{\includegraphics[angle=-0]{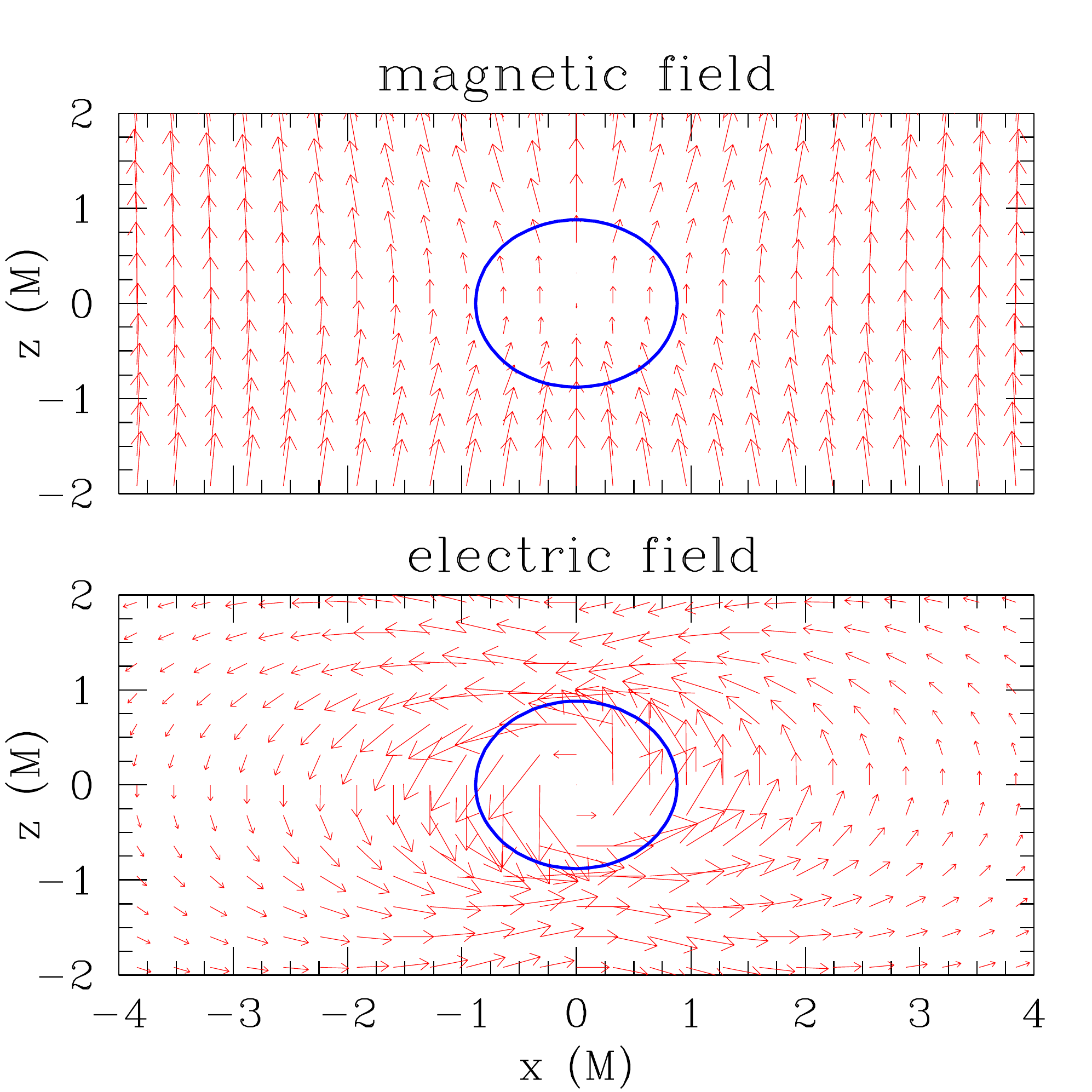}}
  \vskip -0.25cm
  \caption{\textit{Left panel:} Large-scale magnetic and electric
    field lines on the plane $y=0$ and at $t=200\,M$ for a Kerr black
    hole with spin $J/M^2=0.7$ orthogonal to the magnetic field, \ie
    along the $x$-axis. Indicated with blue circles are the apparent
    horizons. \textit{Right panel:} The same as on the left
    panel but on a smaller scale to highlight the fields structure in
    the vicinity of the black hole.}
  \label{fig:fields_na07_y0}
\end{center}
\end{figure*}

In order to obtain a more intuitive picture of the different solutions
for isolated black holes, we now turn our attention to the structure
of the electric and magnetic fields themselves. While those field
lines are gauge-dependent, they can be used to determine the effect of
the spin orientation of the black holes on the
solution. Fig.~\ref{fig:fields_a0} shows therefore the
three-dimensional (3D) EM field configurations at late simulation
times when the solution has settled to a stationary state for either a
Schwarzschild black hole (left panel), or for Kerr black holes with
spin aligned (central panel) or orthogonal to the magnetic field
(right panel). Note that in all of the panels, the magnetic field
lines are bent by the black hole geometry. The appearance of toroidal
electric field in the case of a nonspinning black hole does not
contradict Wald's solution, for which it should be identically
zero. It is due to the non-vanishing radial shift vector which, when
coupled with the vertical magnetic field, leads to a toroidal magnetic
field~\cite{Palenzuela:2009hx}. Finally, note that whenever the black
hole is rotating, together with the gauge-induced toroidal electric
field, there appears also a poloidal component which is induced by the
gravitomagnetism (or frame-dragging) of the rotating black hole and
whose detailed geometry depends on the relative orientation of the
spin with respect to the background magnetic field. For compactness we
do not report here the EM field configuration for a rotating black
hole with spin anti-aligned with respect to the magnetic field. It is
sufficient to remark that the solution shows the same behavior as the
aligned case, with a simple reversal in the direction of the
spin-induced effects.

\begin{figure*}[t]
\begin{center}
   \scalebox{0.4}{\includegraphics[angle=-0]{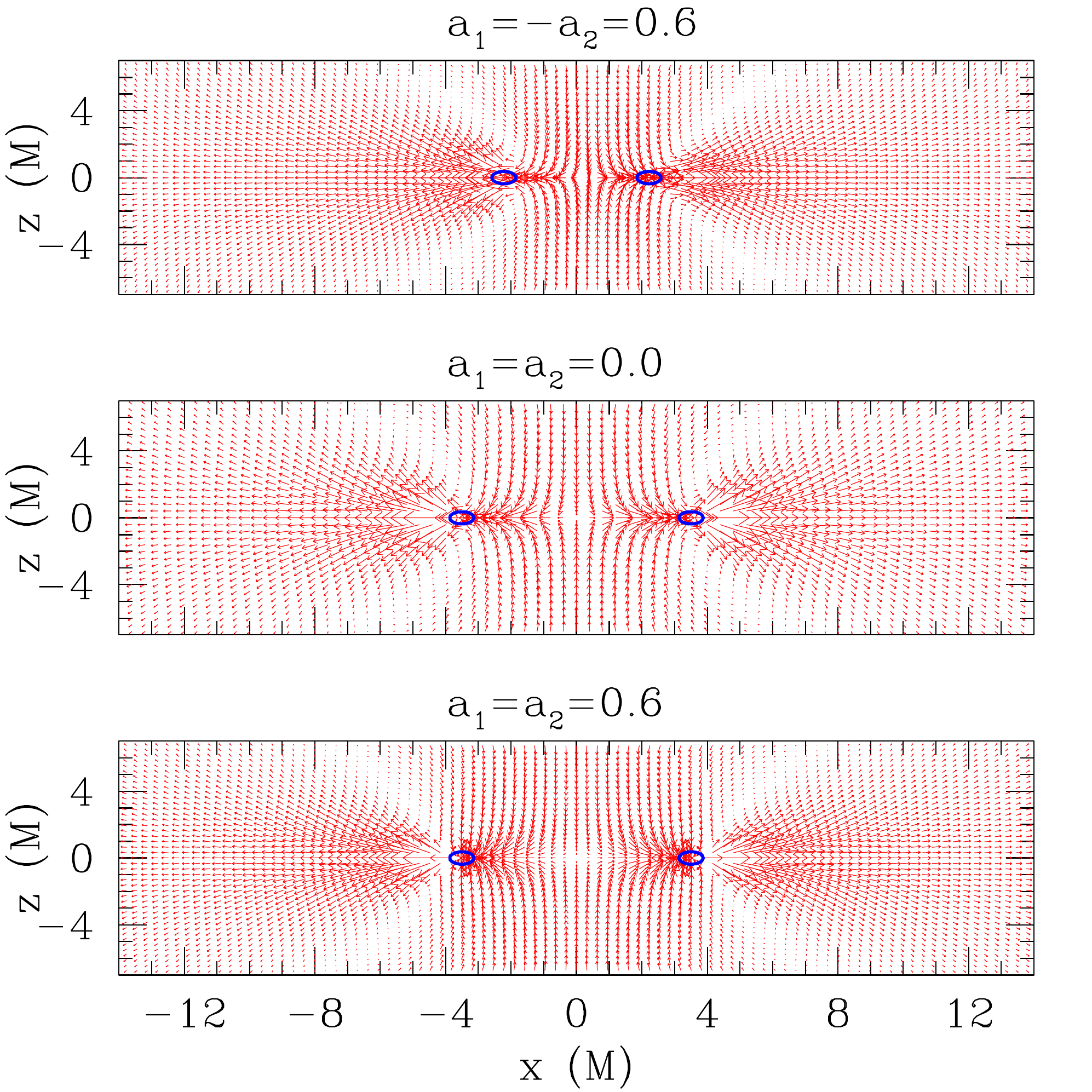}}
  \hskip 1.0cm
   \scalebox{0.4}{\includegraphics[angle=-0]{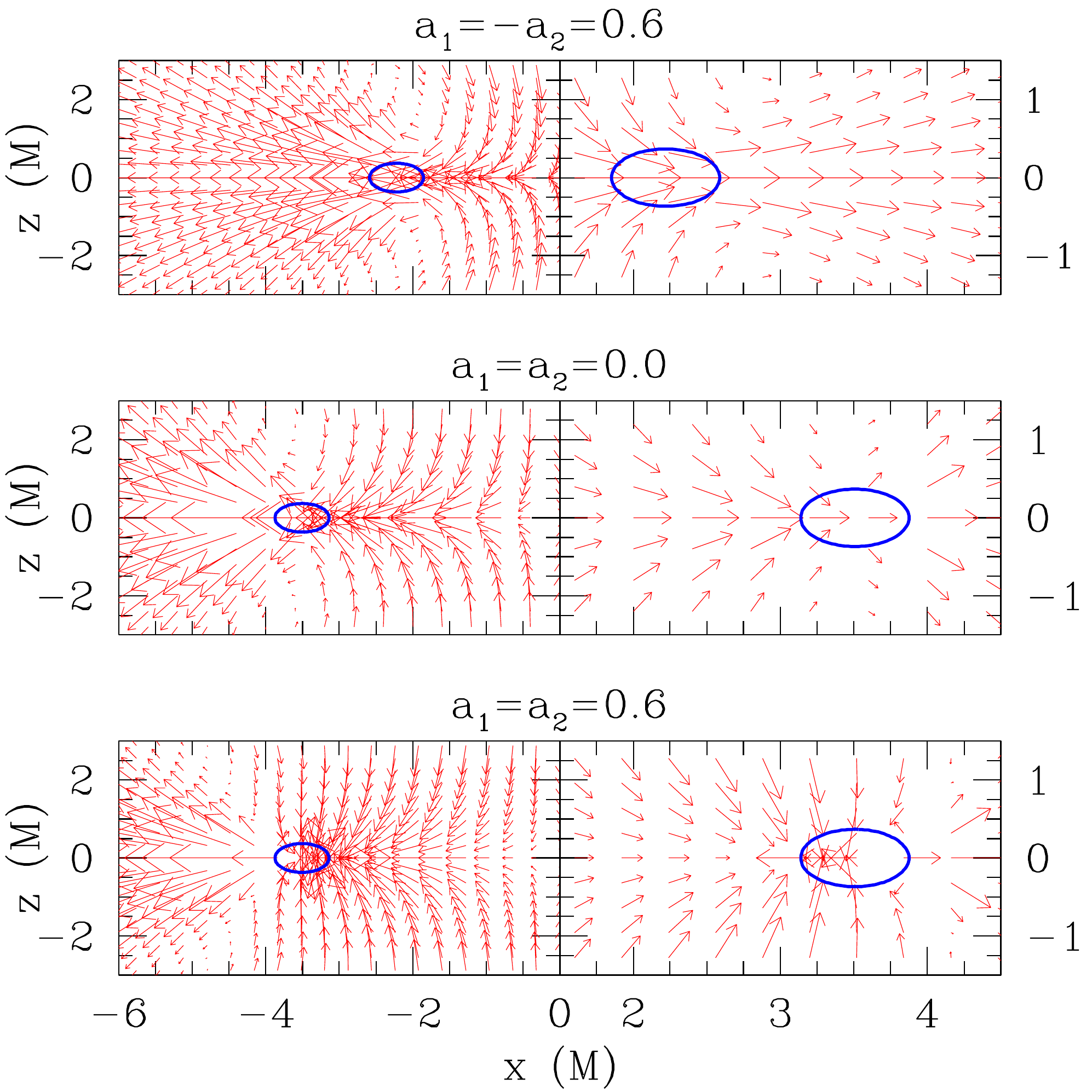}}
  \vskip 0.5cm
  \caption{Electric field lines on the plane $y=0$ for the $r_0$,
    $s_0$ and $s_6$ configurations at $t=123, 155$ and $246\,M$,
    respectively. \textit{Left panel:} Large-scale structure of the EM
    fields around the apparent horizons (blue circles). \textit{Right
      panel:} The same as on the left but on a smaller scale to
    highlight the field structure in the vicinity of the black
    holes. Note that an additional magnification is applied to the
    black hole ``on the right'' so as to highlight the change of sign
    in the electric field near the horizon, \ie at $x\simeq 3\,M$.}
  \label{fig:Elines}
\end{center}
\end{figure*}

To gain some insight on the influence of the black hole spin and
orientation on the EM field lines, it is useful to exploit the
phenomenological description offered by the ``membrane
paradigm''~\cite{Price86a}. In such approach, the horizon of a
rotating black hole is seen as a one-way membrane with a net a surface
charge distribution which, for the case of aligned spin and magnetic
field, has negative values around the poles while positive ones around
the equator. The resulting behavior is therefore the one shown in
Fig.~\ref{fig:fields_a07_y0}, where the magnetic and electric field
lines for the Kerr black hole with spin aligned with the magnetic
field are presented on the $y=0$-plane. The left panel, in particular,
offers a large-scale view of the EM fields, which is however magnified
on the right panels to highlight the behavior of the fields near the
apparent horizons. Finally, shown in Fig.~\ref{fig:fields_na07_y0} are
the magnetic and electric-field lines on the $y=0$-plane for the Kerr
black hole with spin orthogonal to the the magnetic field, \ie along
the $x$-axis. Note that while the differences in the magnetic field
configurations in Fig.~\ref{fig:fields_a07_y0}
and~\ref{fig:fields_na07_y0} are small and difficult to observe even
in the zoomed-in version of the figures, the differences in the
electric fields are instead significant and related to the
different spin orientations.

%
%
\section{Binary black holes}
\label{sec:BBHs}

We next extend the considerations made in the previous section to a
series of black hole binaries having equal masses and spins that are
either aligned or anti-aligned with the orbital angular momentum.

\subsection{Initial Data and Grid Setup}

We construct consistent black-hole initial data via the ``puncture''
method as described in ref.~\cite{Ansorg:2004ds}. We consider equal
mass binaries with four different spin configurations belonging to the
sequences labeled as \textit{``r''} and \textit{``s''} along straight
lines in the $(a_1, a_2)$ parameter space, also referred to as the
``spin diagram''~\cite{Rezzolla:2007xa,Reisswig:2009vc}.  These
configurations allow us to cover the basic combinations for the
alignment of the spin of the individuals black holes with respect to
the magnetic field, while keeping the dimensionless spin parameter of
the single black holes constant among the different binaries
considered. Furthermore, it allows us to study the impact that the
final black hole spin has on the late stages of the merger.

We note that similar sequences have also been considered
in~\cite{Koppitz-etal-2007aa,Pollney:2007ss, Rezzolla-etal-2007,
  Rezzolla-etal-2007b, Rezzolla-etal-2007c} but have here been
recalculated both using a higher resolution and with improved initial
orbital parameters. More specifically, we use post-Newtonian (PN)
evolutions following the scheme outlined in~\cite{Husa:2007rh}, which
provides a straightforward prescription for initial-data parameters
with small initial eccentricity, and which can be interpreted as part
of the process of matching our numerical calculations to the inspiral
described by the PN approximations. The free parameters of the
puncture initial data we fix are: the puncture coordinate locations
${\boldsymbol C}_i$, the puncture bare mass parameters $m_i$, the
linear momenta ${\boldsymbol p}_i$, and the individual spins
${\boldsymbol S}_i$.  The initial parameters for all of the binaries
considered are collected in Table~\ref{tableone}. The initial
separations are fixed at $D=8\,M$ with the exception of the $s_{-6}$
binary having an initial separation of $D=10\,M$.  Here $M$ is the
total initial black hole mass, chosen as $M=1$ (note that the initial
ADM mass of the spacetime is not exactly $1$ due to the binding energy
of the black holes), while the individual asymptotic initial black
hole masses are therefore $M_i = 1/2$. In addition, we choose the
initial parameters for the EM fields to be $B^i = (0,0,B_0)$ with $B_0
\sim 10^{-4}/M\sim 10^8(10^8 M_{\odot}/M)\,$G and $E^i = 0$. The setup
for the numerical grids used in the simulations consists of 9 levels
of mesh refinement with a fine-grid resolution of $\Delta x/M=0.02$
together with fourth-order finite differencing. The wave-zone grid has
a resolution of $\Delta x/M=0.128$ and extends from $r=24\,M$ to
$r=180\,M$, in which our wave extraction is carried out.  The outer
(coarsest) grid extends to a spatial position which is $819.2\,M$ in
each coordinate direction.

\begin{figure*}[t]
\begin{center}
   \scalebox{0.4}{\includegraphics[angle=-0]{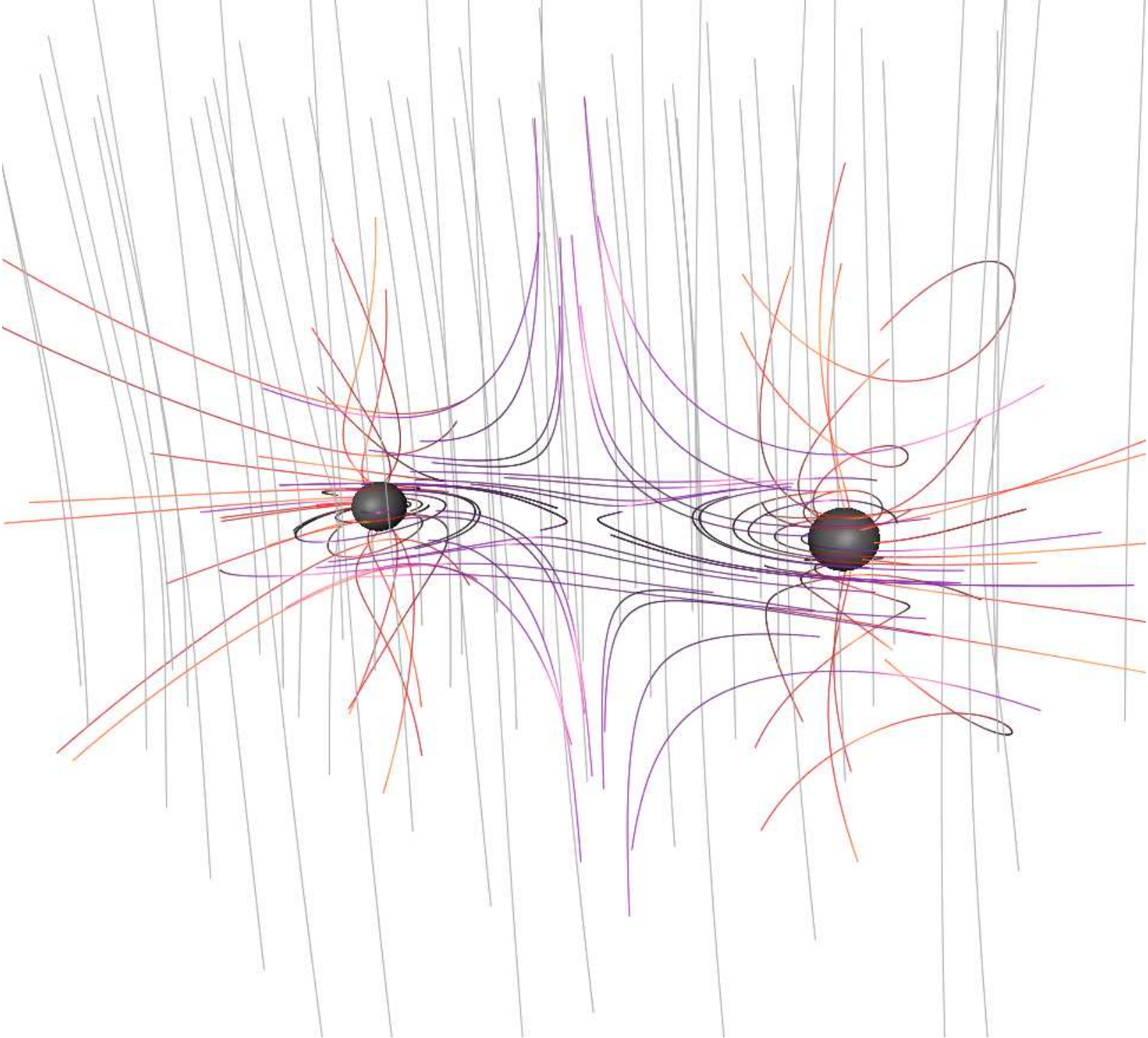}}
   \scalebox{0.5}{\includegraphics[angle=-0]{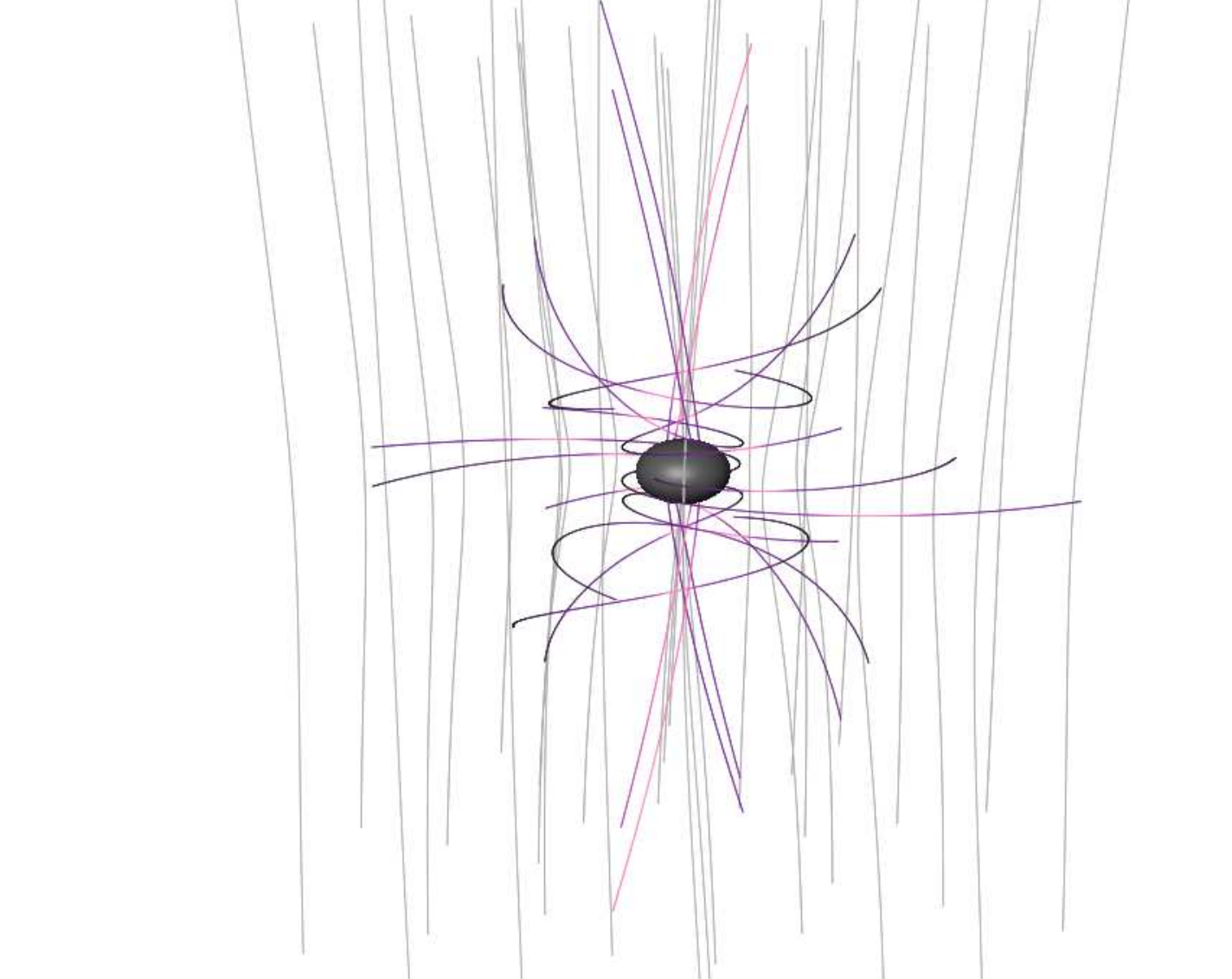}}
   \vskip 0.5cm
  \caption{Electric (red/magenta) and magnetic field lines (gray) in
    3D for the $s_6$ binary during inspiral when both black holes are
    still far separated at time $t=328\,M$ (left panel), and after the
    merger at $t=690\,M$ (right panel).}
  \label{fig:Efields}
\end{center}
\end{figure*}

\begin{table*}
\caption{\label{tableone}Binary sequences for which numerical
  simulations have been carried out, with various columns referring to
  the puncture initial location $\pm x/M$, the mass parameters
  $m_i/M$, the dimensionless spins $a_i$, the initial momenta and the
  normalized ADM mass ${\widetilde M_{_{\rm ADM}}}\equiv M_{_{\rm
      ADM}}/M$ measured at infinity. (See
  refs.~\cite{Rezzolla:2007xa,Reisswig:2009vc} for a discussion of the
  naming convention).}
\vspace{0.1cm}
\begin{ruledtabular}
\begin{tabular}{|l|ccccccc|}
~					&
{$\pm x/M$} 				&
{$m_1/M$} 				&
{$m_2/M$} 				& 
{$a_1$} 				&
{$a_2$} 				&
{$(p_x,~p_y)_1=-(p_x,~p_y)_2$}           &
{${\widetilde M_{_{\rm ADM}}}$} 	\\
\hline
$s_{-6}$   & $5.0000$ & $0.4000$ & $0.4000$ & $-0.600$      & $-0.600$      & $(0.001191,-0.100205)$ & $0.9873$ \\
$r_0$   & $4.0000$ & $0.4000$ & $0.4000$ & $-0.600$      & $\pls 0.600$  & $(0.001860,-0.107537)$ & $0.9865$ \\
$s_0$   & $4.0000$ & $0.4824$ & $0.4824$ & $\pls 0.000 $ & $\pls 0.000 $ & $(0.002088,-0.112349)$ & $0.9877$ \\
$s_6$   & $4.0000$ & $0.4000$ & $0.4000$ & $\pls 0.600 $ & $\pls 0.600 $ & $(0.001860,-0.107537)$ & $0.9876$ \\
\end{tabular} 
\end{ruledtabular}
\end{table*}

\subsection{Binary Evolution and Spin Dependence}

\begin{figure*}[t]
\begin{center}
   \scalebox{0.4}{\includegraphics[angle=-0]{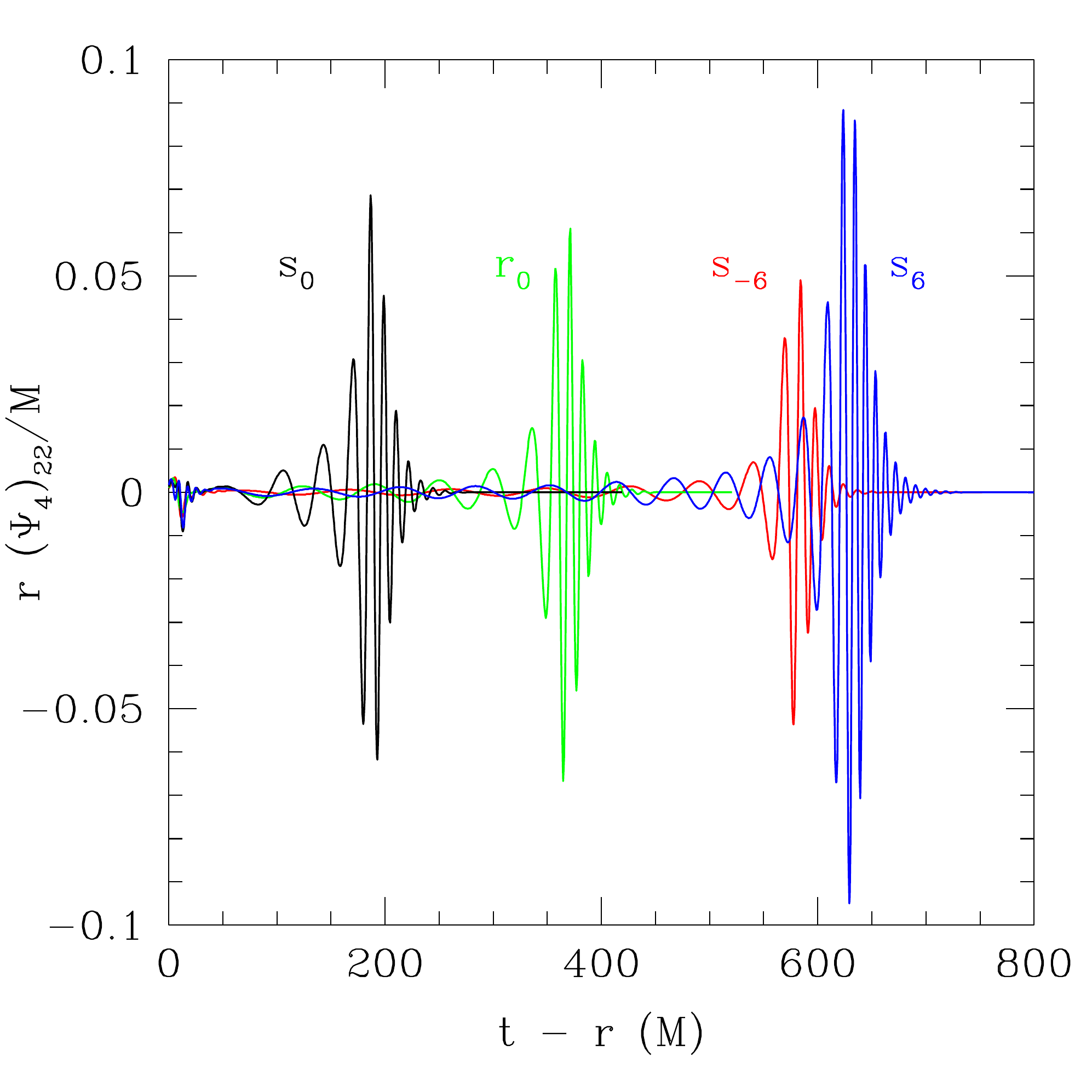}}
   \hskip 1.0cm
   \scalebox{0.4}{\includegraphics[angle=-0]{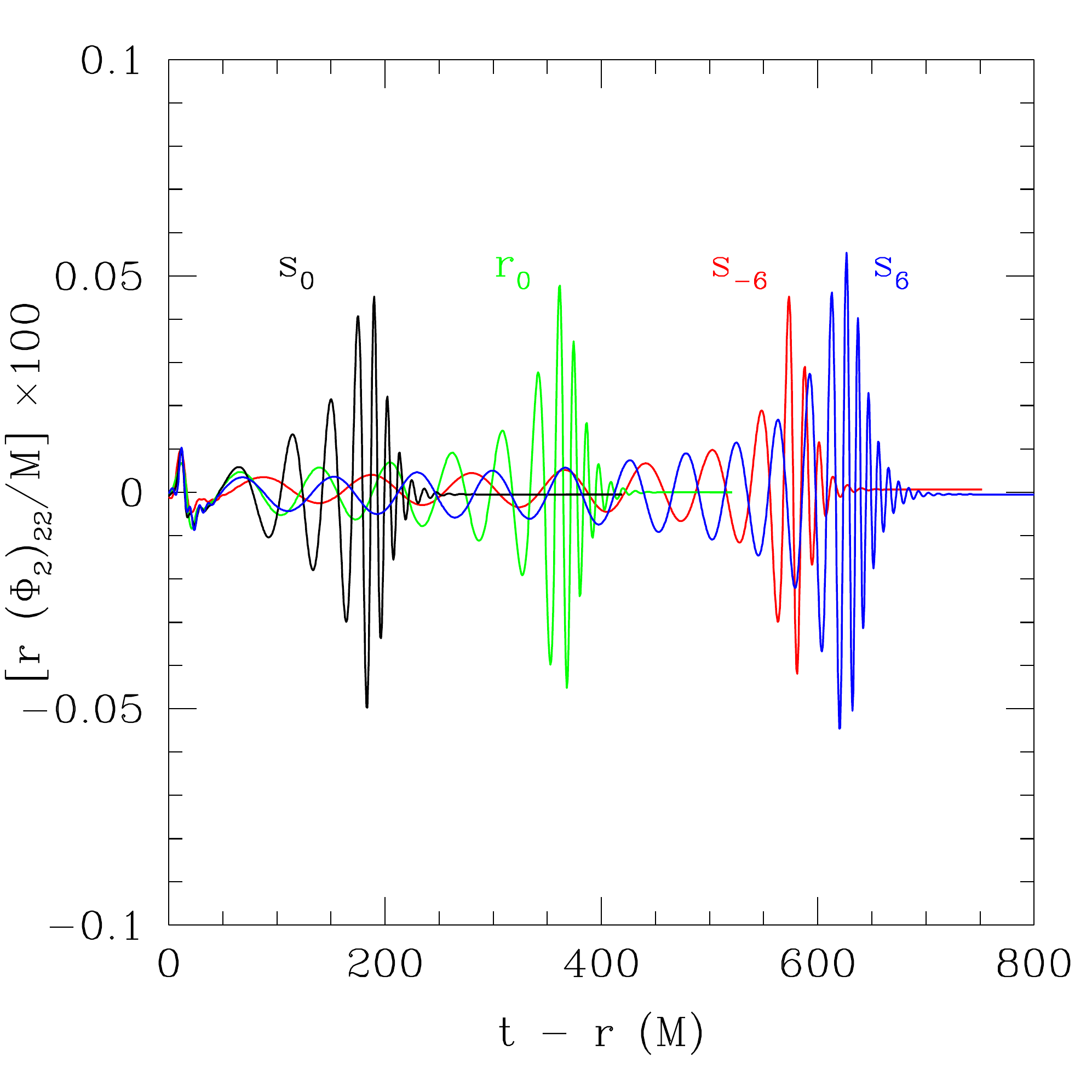}}
  \vskip 0.5cm
  \caption{\textit{Left panel:} GWs as computed from the $(2,2)$ mode
    of $\Psi_4$ for the different binaries reported in Table
    \ref{tableone}. \textit{Right panel:} The same as in the left
    panel but for the EM waveform as computed from
    $\Phi_2$.}
  \label{fig:psiphi22}
\end{center}
\end{figure*}

As mentioned above, we consider configurations where both black holes
have equal mass and the individual black hole spins are either aligned
or anti-aligned with the magnetic field (and orbital angular
momentum). We thus consider a set of three different spinning
binaries, as well as a nonspinning binary, which we take as a
reference (\cf Table~\ref{tableone}).
 
\begin{figure*}[t]
\begin{center}
   \scalebox{0.3125}{\includegraphics[angle=-0]{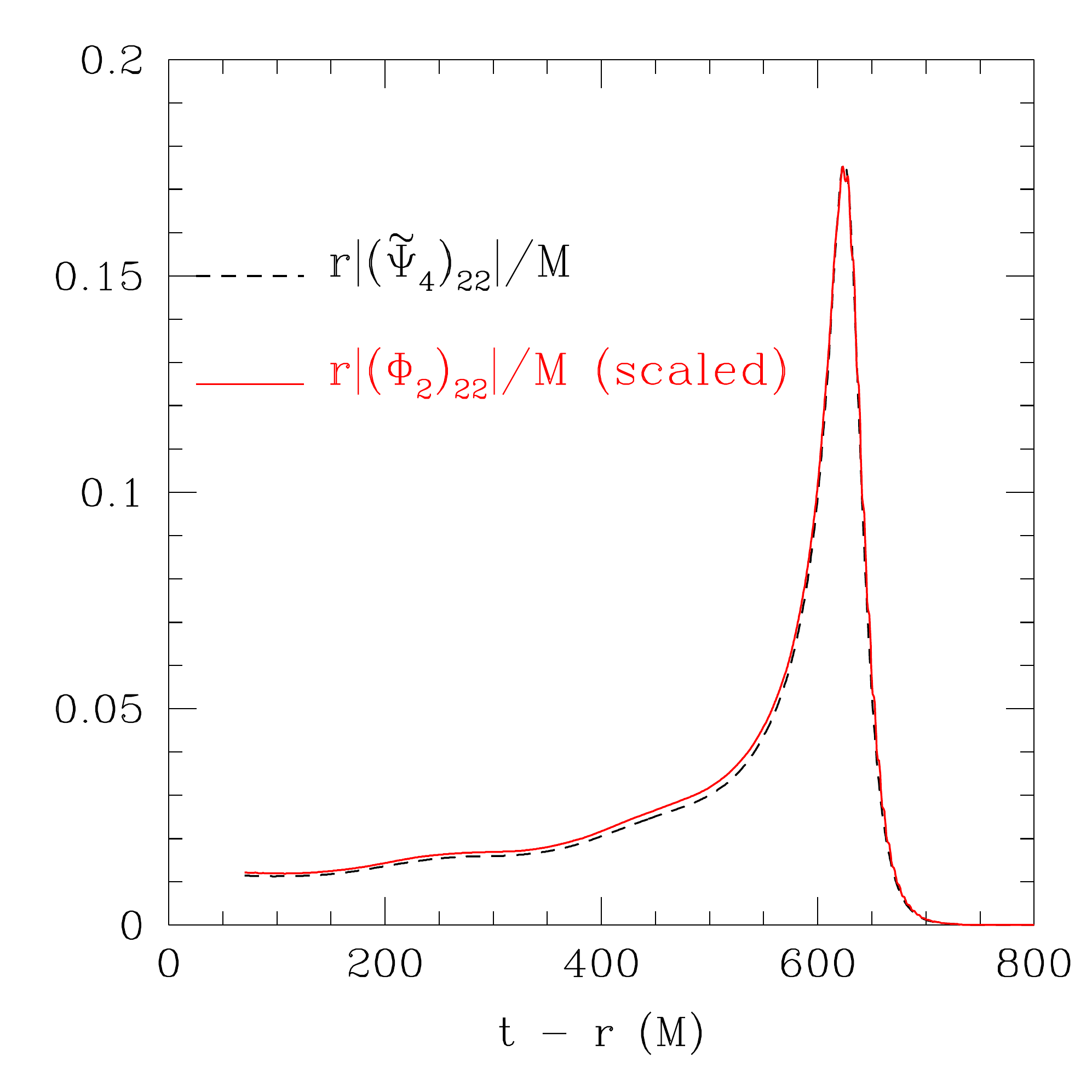}}
   \hskip 2.0cm
   \scalebox{0.3125}{\includegraphics[angle=-0]{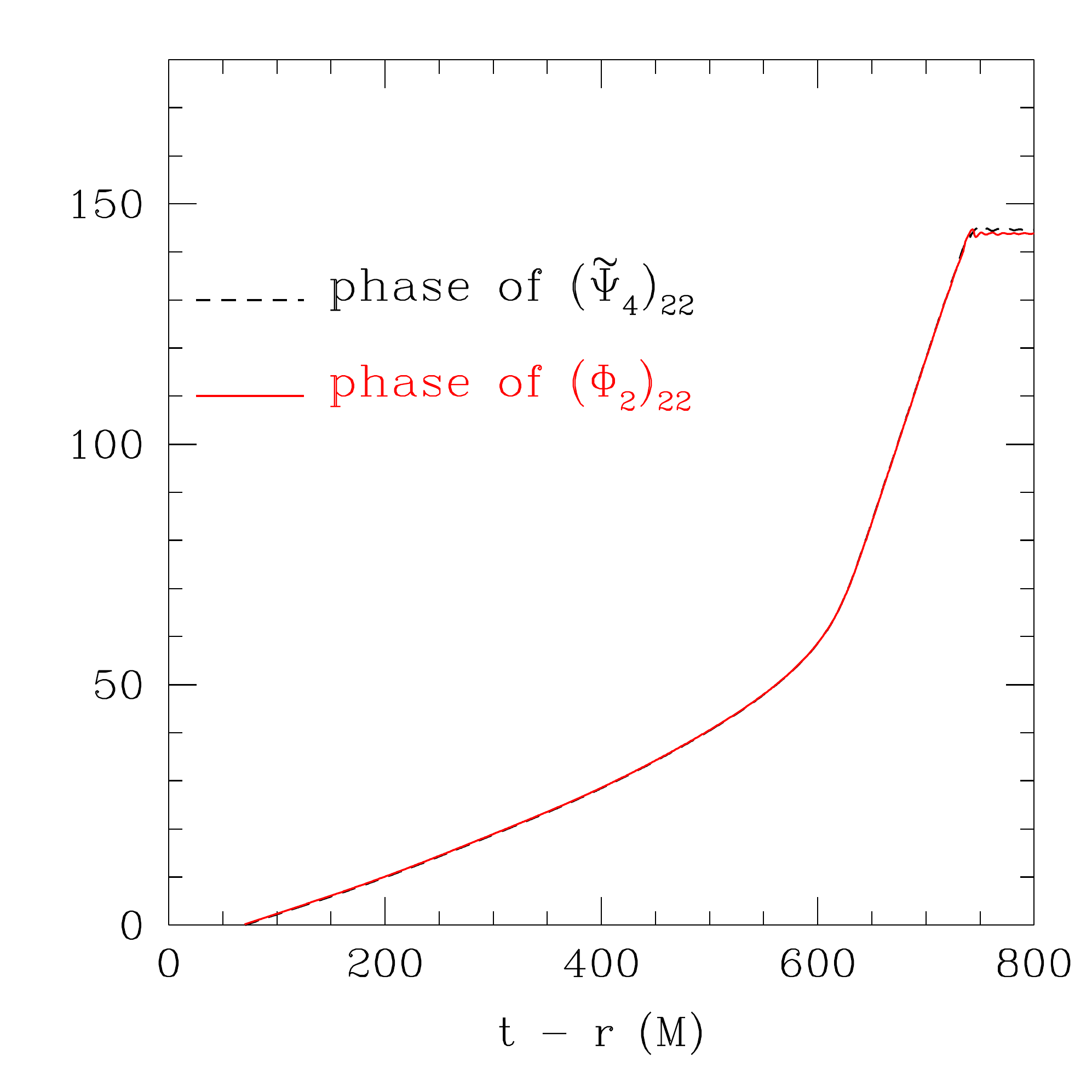}}
   \scalebox{0.3125}{\includegraphics[angle=-0]{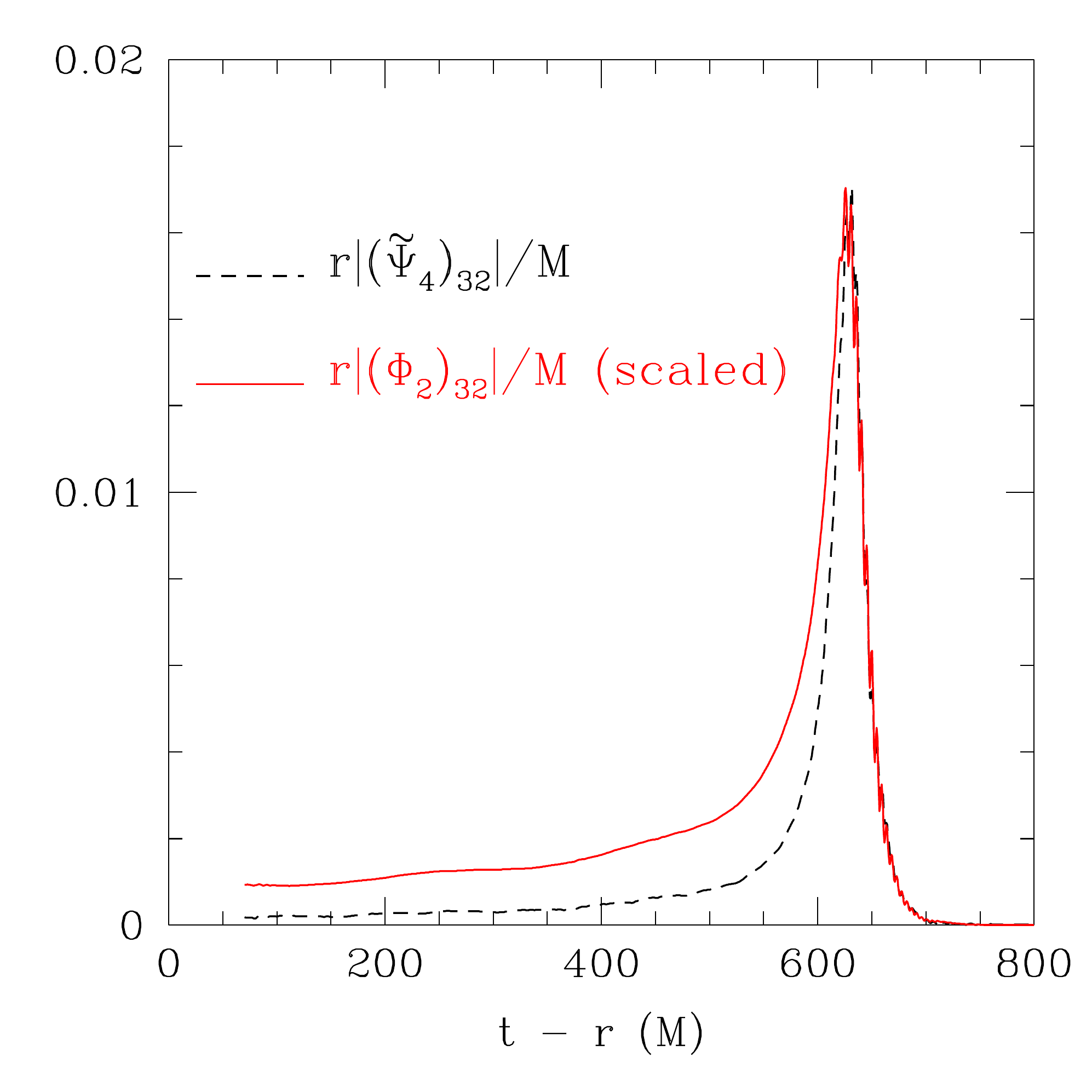}}
   \hskip 2.0cm
   \scalebox{0.3125}{\includegraphics[angle=-0]{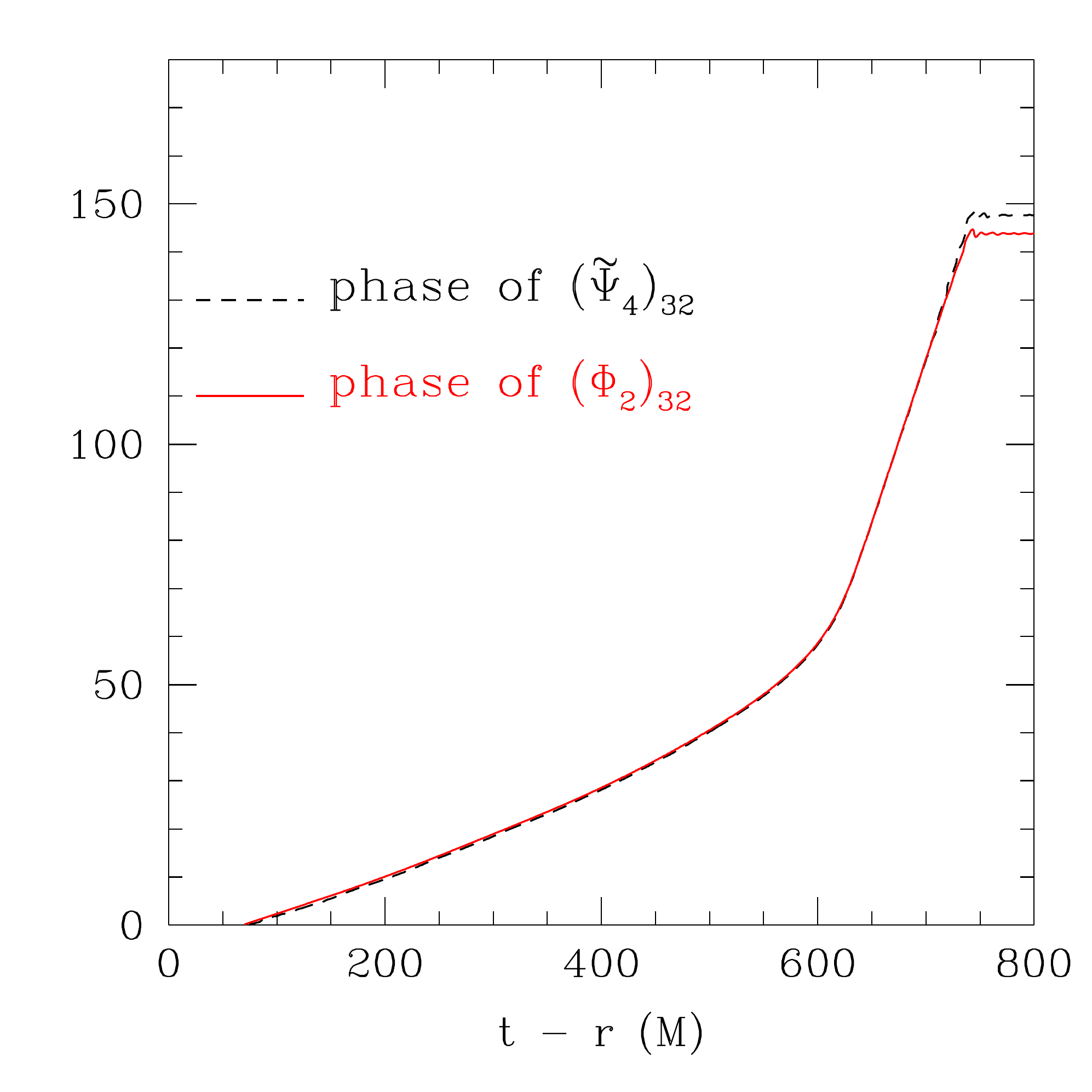}}
  \vskip -0.125cm
  \caption{Amplitude and phase evolution of the main $\ell=m=2$ modes
    for the Weyl scalar $\Phi_2$ and the first time integral of
    $\Psi_4$ (\ie $\widetilde{\Psi}_4$), relative to the $s_6$
    configuration. The plots show the data in retarded time $t-r$ for
    a detector located at $r=100M$. While the $\ell=m=2$ modes show
    the same amplitude (up to a scale factor) and phase evolution,
    this does not apply to modes with higher $\ell$. For the
    $\ell=3,m=2$ and $\ell=4,m=2$ modes, the phase evolution
    is still identical but the amplitude no longer does not differ 
    only by a constant scale factor.}
  \label{fig:amp_phase_uu}
\end{center}
\end{figure*}

One feature of our simulations, that was already analysed for single
black holes in Sect.~\ref{sec:single_BHs}, and is of even greater
interest for binaries, is the structure of the EM field lines induced
by the spacetime dynamics around the black holes. The field line
configurations, in fact, change considerably throughout the course of
our simulations.  When there is a large separation between the
orbiting black holes, the electric field structure in both nonspinning
and spinning binary systems is dominated by the orbital motion of the
individual black holes. In particular, an inspection of the electric
field vector along a line joining their centers indicates an outward
radial dependence. This can be understood from the phenomenological
interpretation suggested by the membrane paradigm and has been
observed already in~\cite{Palenzuela:2009hx}. Namely, as the black
holes move in a direction which is essentially orthogonal to the
magnetic field, an effective quadrupolar charge separation develops on
the horizons with effective positive charges at the poles and negative
ones on the equator, thus inducing an electric field emanating from
each black hole. This induced quadrupolar electric field is therefore
reminiscent of the one produced by a conductor moving through a
uniform magnetic field as the result of the Hall effect.

It is interesting to note that while the differences in the magnetic
field lines among the various binaries considered are rather small,
the differences in the electric fields show significant variation
across the spin configurations. This is
illustrated in Fig.~\ref{fig:Elines}, which shows the electric field
lines at different scales of interest with respect to the black holes
for two spinning binary black hole systems and the nonspinning binary
on the $y=0$-plane. Here we choose to concentrate on the
configurations with the spins up/up (\ie $s_6$) and with the spin
up/down (\ie $r_0$) since the configuration with the spins down/down
(\ie $s_{-6}$) shows the same field-line structure as the up/up case.
In particular, the left panel of Fig.~\ref{fig:Elines} reports the
field-line structure on a scale which is much larger than that of the
horizons and that clearly shows the quadrupolar nature of the
field. At the same time, the right panel offers a magnified view of
the same binaries on scales which is comparable with those of the
horizons. In this way it is possible to find the properties of the 
electric field already discussed in Sect~\ref{sec:single_BHs} for 
isolated black holes also in case of binary black holes. Additionally the
various spin configurations lead to different small-scale properties
of the field. More specifically, while the field lines of the $r_0$
and $s_0$ configurations have a similar structure even in the
magnified plot, the binary with the aligned spins $s_6$ shows a more
complex structure in which the electric field changes sign near but
outside the horizon, namely at $x\simeq 3\,M$ and which corresponds
approximately to a distance $d\sim 2 r_{_{\rm AH}}$, with $r_{_{\rm
    AH}}$ the mean radius of the apparent horizon. This additional
property of the electric field could be related to the location of the
ergosphere (which has not been computed in these simulations) and may
be seen as a response of the electric field to the additional charge
separation induced on the black hole horizon and which leads to a
greater distortion and twisting of the field lines in this region.

Although it is not trivial to disentangle how much of this behaviour
of the electric field depends on the gauges used, the complex structure of
the electric fields, and which varies considerably through the late
inspiral and the merger of the binary, may lead to interesting
dynamics and to the extraction of energy via acceleration of particles
along open magnetic field lines or via magnetic reconnection.  To
better illustrate the complex field structure, Fig.~\ref{fig:Efields}
offers 3D views of both the electric and magnetic fields for the $s_6$
binary. In particular, in the left panel of Fig.~\ref{fig:Efields} we
show the electric and magnetic field lines as well as the apparent
horizons when the binary is inspiralling (\ie at $t=328\,M$) and again
observe the superposition of two effects: the overall orbital motion
of the black holes causing the large scale structure of the electric
field lines (highlighted in a magenta color); and the effect of the
black hole spin (in red), which causes additional dragging in the
electric field lines close to the apparent horizons. In the right
panel, on the other hand, we present the late-time (\ie at $t=690\,M$)
state of the solution which, as expected, agrees well with the field
line configurations presented for the Kerr black hole with spin
aligned with the magnetic field in Sect.~\ref{sec:single_BHs}.

\begin{figure*}[t]
\begin{center}
   \scalebox{0.4}{\includegraphics[angle=-0]{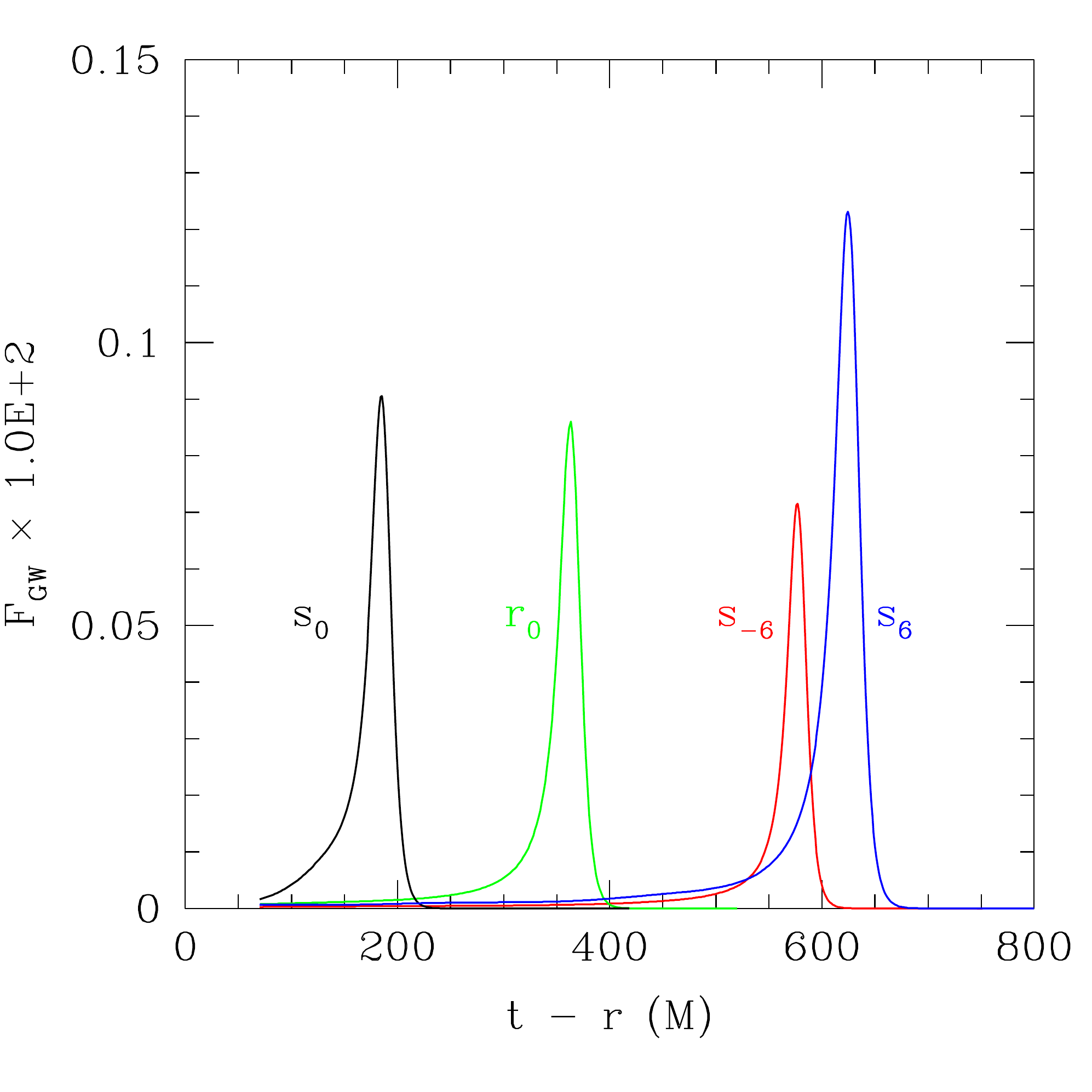}}
   \hskip 1.0cm
   \scalebox{0.4}{\includegraphics[angle=-0]{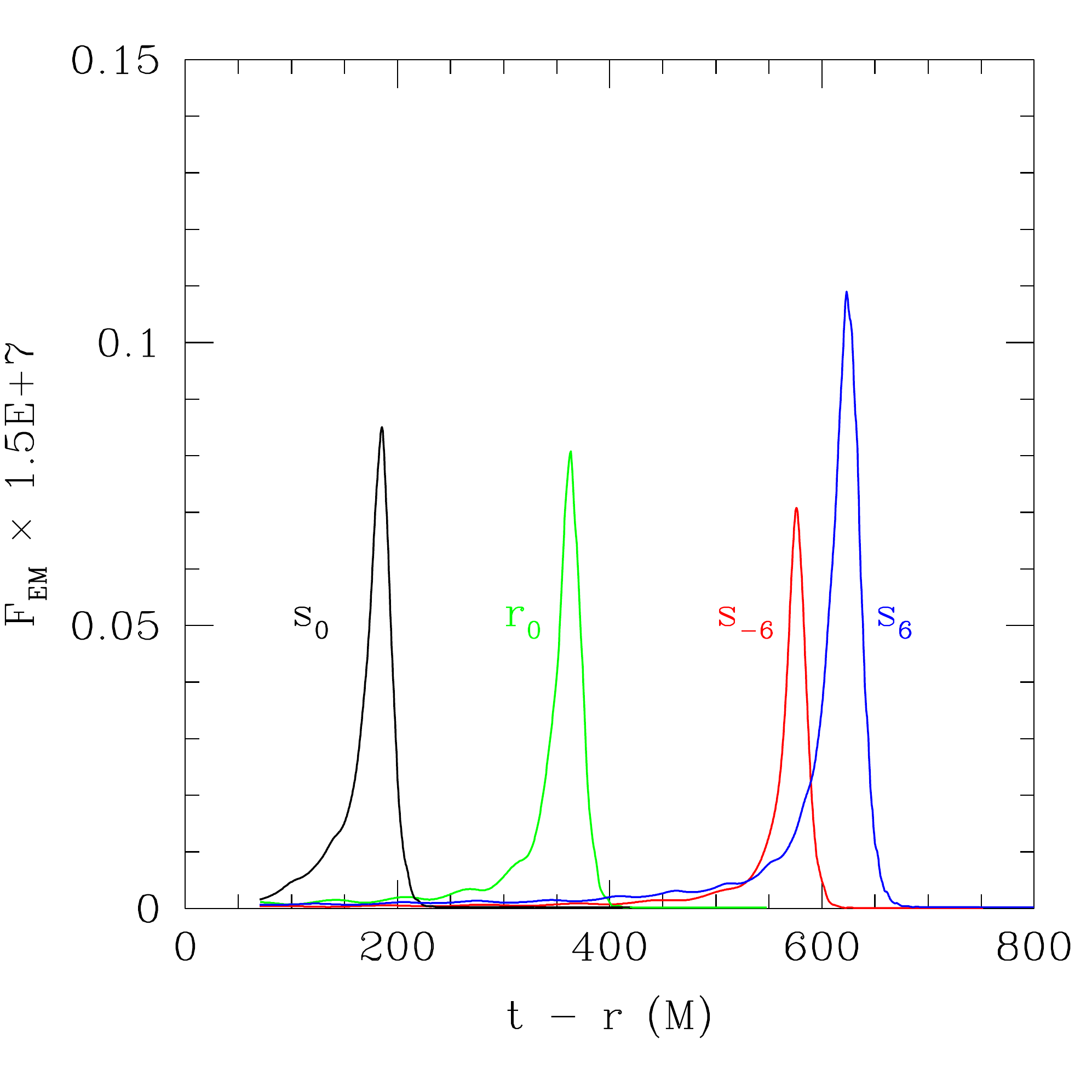}}
  \vskip 0.5cm
  \caption{The total energy flux per unit solid angle in terms of GW
    waves (left panel) and of EM waves (right panel); clearly they
    differ only up to a scaling factor. The different lines refer to
    the different binaries reported in Table \ref{tableone}.}
  \label{fig:radiatedE}
\end{center}
\end{figure*}

We next switch our attention to discussing how the different BH spin
configurations affect the emission of EM radiation. This requires a
careful analysis of the radiative properties of the solution in both
the EM and gravitational channels.  We first focus on the two types of
waveforms and Fig.~\ref{fig:psiphi22} illustrates the correlation
between the two emissions by showing the time-retarded waveform of the
principal mode, \ie the $\ell=m=2$ of the spin-weighted spherical
harmonic basis (note that $\Psi_4$ and $\Phi_2$ have spin weight
$-2$ and $-1$, respectively), for all different spin configurations to 
compare gravitational and EM waveforms directly. While both EM and GW
radiation show the same characteristics in the dominant mode, we note
that small differences arise when comparing the waveforms of the
individual spin configurations more carefully with each other in the
two channels. Note that the waveform for the binary $s_{-6}$ has a
larger number of cycles only because it merges very rapidly (the total
angular momentum is smaller because the total spin is anti-parallel to
the orbital angular momentum) and thus it has been evolved from a
larger initial separation $D=10\,M$; all the other binaries have the
same initial separation $D=8\,M$.  A closer inspection of
Fig.~\ref{fig:psiphi22} reveals that the amplitude evolution of the
$\ell=m=2$-mode for the different spin configurations differs when
compared in the two channels. As an example, while in the GW channel
the amplitude in the $\ell=2,m=2$-mode decreases when going from the
$r_0$-configuration over to the $s_0$ and $s_{-6}$ configurations, the
amplitude remains nearly constant in the EM channel. This reveals that
there are additional contributions in the EM emission coming from
the higher-order modes (see Fig.~\ref{fig:radiatedE} and the
discussion below)

To further evaluate the correlation between the EM and the
gravitational radiation, we now turn our attention to the amplitude
and phase evolution of the main contributing spherical harmonic
modes. Since radiated energy fluxes are given by $\Phi_2$ and the time
integral of $\Psi_4$ we here compare $\Phi_2$ with
$\widetilde{\Psi}_4\equiv \int_{\infty}^t \Psi_4 dt'$. For briefness
we only highlight the results obtained for the $s_6$ configuration,
since this shows the highest amount of energy being radiated in both
EM and GWs, and because our remarks apply also to the other
configurations. Since the main contributions to the radiated energy in
the EM channel arise from the $\ell=2,3,4,\ m=2$-modes, we limit our
analysis in this section to those modes only. In order to obtain a
better understanding of the correlation in the radiation coming from
the two channels, we analyse the amplitude and phase of the main
contributing modes individually.  Fig.~\ref{fig:amp_phase_uu} shows
the amplitude and phase evolution of the $\ell=2,3,\ m=2$-modes 
in both channels. Clearly, the $\ell=m=2$ modes
show the same phase evolution (\cf the left panels) in the two forms
of radiation, as expected given that the EM emission is essentially
driven by the orbital motion of the binary.  Furthermore, the
amplitude evolution in the $\ell=2,m=2$ modes of both emissions are
also simply related by a constant, time-independent, factor.

Although a simple scaling factor in the evolution of
$\widetilde{\Psi_4}$ and $\Phi_2$ appears for all of the different
binary configurations considered here, this factor is not the same
across different spin configurations. However, because the $\ell=m=2$
represents by and large the most important contribution to the
radiation emitted in the two channels and because the GW-emission from
binaries with spins aligned/antialigned with the orbital angular
momentum has been computed in a number of related
works~\cite{Campanelli:2006uy, Campanelli:2006vp, Campanelli:2006gf,
  Buonanno:07b, Koppitz-etal-2007aa, Herrmann:2007ex, Herrmann:2007ac,
  Pollney:2007ss, Rezzolla-etal-2007, BoyleKesdenNissanke:07,
  BoyleKesden:07, Marronetti:2007wz,Rezzolla-etal-2007b,
  Rezzolla-etal-2007c,Barausse:2009uz}, the results found here allow
us to simply extend all of the phenomenology reported so far for the
$\ell=2,m=2$ GW-emission from the above cited works also to the EM
channel.

Unfortunately, the tight correlation found in the amplitude evolution
of the lowest-order mode disappears for higher-order modes. This is
reported in the bottom panels of Fig.~\ref{fig:amp_phase_uu}, which
indicate that while the phase amplitude remains the same (\cf the
bottom right panel), the evolution of the amplitude in the two
channels does not differ only by a simple constant scaling factor (\cf
the bottom left panel). A similar behaviour is found for lower-order
modes such as the $\ell=4,m=2$ one but is not reported here for
compactness. Overall, these results suggest that although the main
(and lowest-order) contribution to the EM emission does indeed come as
a result of the dragging of the EM fields by the orbital motion of the
binary, additional contributions arise from higher-order modes 
which are not directly related to the orbital motions. These are
likely to be the result of the complex interactions among the EM
fields, discussed in Fig.~\ref{fig:Elines} and whose investigation,
although of great interest, goes beyond the scope of this paper.

Another interesting quantity to consider in our analysis is the energy
carried away from the systems in the two emissions, which can be
computed by using eqs.~(\ref{FGW_modes})--(\ref{FEM_modes}), where we
have taken into account modes up to $\ell=4$. Despite the differences
between the EM and gravitational waves discussed already, when looking 
at the emission in the lowest-order modes that can be associated 
to the different multipolar decomposition of the two emissions (\cf
Fig.~\ref{fig:psiphi22}), we find that the overall energy fluxes are
extremely similar and differ essentially only by a constant (but
large) factor. This is shown in Fig.~\ref{fig:radiatedE} which reports
both the GW (left panel) and the EM radiated energy fluxes (right
panel) when integrated over a sphere located at $r=100\,M$ for all
the binary sequences considered here. Once again, the fact that
$F_{_{\rm EM}}$ basically mimics $F_{_{\rm GW}}$, underlines that the
emission in the EM channel is intimately tied to the emission in GW,
so that the observation of one of the two would lead to interesting
information also about the other one. As a final comment it is worth noting
that although the energy fluxes from the binaries $s_0$ and $r_0$ show
a different evolution, the total emitted energy, namely the area under
the curves, is extremely similar and is reported in
Table~\ref{tabletwo}. This provides yet an additional confirmation of 
the results already presented in
refs~\cite{Rezzolla-etal-2007,Rezzolla-etal-2007b,
  Rezzolla-etal-2007c,Reisswig:2009vc,Barausse:2009uz} for binaries
with aligned spins and yields further support to the conjecture that
when the initial spin vectors are equal and opposite and the masses
are equal, the overall dynamics of the binary is the same that of the
corresponding nonspinning binary.

%
%
\section{Astrophysical Detectability}
\label{sec:astrophysics}

\begin{table}
\caption{\label{tabletwo}Relative emitted energies in EM waves and GWs
($E^{\rm rad}_{_{\rm EM}}/M$, $E^{\rm rad}_{_{\rm GW}}/M$, 
respectively), and emitted angular momentum in GWs ($J^{\rm rad}_{_{\rm GW}}/M^2$),
for the magnetic field $B_0\,M = 10^{-4}$.}
\vspace{0.1cm}
\begin{ruledtabular}
\begin{tabular}{|l|ccc|}
model				&
$E^{\rm rad}_{_{\rm EM}}/M$	&
$E^{\rm rad}_{_{\rm GW}}/M$		&
$J^{\rm rad}_{_{\rm GW}}/M^2$	\\ 
\hline
$s_{-6}$  & $1.562{\rm E}-7$ & $0.0243$ & $0.216$\\
$r_0$    & $2.040{\rm E}-7$ & $0.0357$ & $0.213$\\
$s_0$    & $2.055{\rm E}-7$ & $0.0354$ & $0.243$\\
$s_6$    & $3.412{\rm E}-7$ & $0.0590$ & $0.380$\\
\end{tabular} 
\end{ruledtabular}
\end{table}

As discussed in the previous sections, the EM and GW radiation are
tightly coupled and evolve on exactly the same timescales and with the
same spectral distribution in frequency. The \textit{rates} of loss of
energy and angular momentum, however, are very different. This is
summarized in Table~\ref{tabletwo} which reports the total energy
radiated during the inspiral and merger in either EM waves or GWs (\ie
$E^{\rm rad}_{_{\rm EM}}/M$, $E^{\rm rad}_{_{\rm GW}}/M$) and the
angular momenta radiated in GWs (\ie $J^{\rm rad}_{_{\rm GW}}/M^2$).
From the values obtained, two interesting observations can be
made. The first one is that the radiated EM energy is higher for
binaries which lead to a more highly spinning final black hole. This
is a consequence of these binaries merging with increasingly tighter
orbits and at higher frequencies, which leads to stronger EM and GW
fluxes. The second one has already been mentioned in the previous
Section and reflects the fact that the binaries $r_0$ and $s_0$ lead
to the same energy emission (and to the same final black-hole
spin~\cite{Rezzolla-etal-2007,Rezzolla-etal-2007b,
  Rezzolla-etal-2007c,Barausse:2009uz}) despite the $s_0$ binary has
black holes with non-zero individual spins.

Note also that, in contrast with the losses in the GW emission, those
in the EM one do not depend just on the masses and initial spins of
the black holes but also on the strength of the initial magnetic
field. This dependence must naturally scale quadratically with the
magnetic field, so that we can write
\begin{eqnarray}
\frac{E^{\rm rad}_{_{\rm EM}}}{M} &=& k_1(a_1,a_2, M_1, M_2) B^2_0 \\
& = & 1.43\times 10^{-32} k_1 \left(\frac{M}{M_{\odot}}\right)^2 
\left(\frac{B}{1\ {\rm G}}\right)^2\,,\qquad 
\end{eqnarray} 
where we have used the following relation 
\begin{equation}
B\ [{\rm G}] = 8.36\times 10^{19} \left(\frac{M_{\odot}}{M}\right) 
B\ [{\rm geom.\ units}]\,.
\end{equation}
to convert a magnetic field in geometric units ($B\ [{\rm
    geom.\ units}]$) into a magnetic field expressed in Gauss
($B\ [{\rm G}]$).

As discussed before, the EM emission is closely related via
simple scaling factors to the GW one and whose efficiency has been
discussed in detail in Sect. VB of ref.~~\cite{Reisswig:2009vc}. In
particular, it was shown there that the radiated GW energy depends
quadratically on the total dimensionless spin (see eq. (24)
in~\cite{Reisswig:2009vc}) and the corresponding coefficients ${\tilde
  p}_i$ were presented in eq.~(25) in the same reference. Hence, at
least in the case of equal-mass binaries, it is trivial to express
$k_1(a_1,a_2, M_1, M_2)$ in terms of the suitably rescaled
coefficients ${\tilde p}_i$ in~\cite{Reisswig:2009vc}. Here, however,
because we are interested in much simpler order-of-magnitude
estimates, we will neglect the dependence of $k_1$ on the spins and
simply assume that $k_1\sim 10^{-7}$, so that
\begin{equation}
\frac{E^{\rm rad}_{_{\rm EM}}}{M} \simeq 10^{-15}
\left(\frac{M}{10^8\ M_{\odot}}\right)^2 
\left(\frac{B}{10^4\ {\rm G}}\right)^2\,,
\end{equation} 
where we have considered a total black hole mass of $10^8\msun$ and a
magnetic field of $10^4$ G as representative of the one possibly
produced at the inner edges of the circumbinary
disc~\cite{Milosavljevic05} (see~\cite{Silantev:2009} for a recent
discussion on the strength of magnetic fields in active galactic
nuclei (AGN)).

It should be noted that only when an extremely strong magnetic field of $\sim
10^{11}\,{\rm G}$ is considered, does the EM efficiency become as
large as ${E^{\rm rad}_{_{\rm EM}}}/{M} \simeq 10^{-1}$ and thus
comparable with the GW one. For more realistic magnetic fields,
however, and assuming for simplicity that ${E^{\rm rad}_{_{\rm
      GW}}}/{M} \sim 10^{-2}$ for all possible spins, the ratio of the
two losses is
\begin{equation}
\frac{E^{\rm rad}_{_{\rm GW}}}{E^{\rm rad}_{_{\rm GW}}} \simeq 10^{-13} 
\left(\frac{M}{10^8\ M_{\odot}}\right)^2
\left(\frac{B}{10^4\ {\rm G}}\right)^2\,.
\end{equation}
That is, for a realist value of the initial magnetic field, the GW
emission is $13$ orders of magnitude more efficient than the EM one.
More importantly, however, the frequency of variation
of the EM fields is of the order
\begin{equation}
f_{_{\rm B}} \simeq
(40\,M)^{-1} \simeq 10^{-4} \left(\frac{10^8\msun}{M}\right)\,\hz 
\end{equation}
and therefore much lower than what is accessible via astronomical
radio observations, which are lower-banded to frequencies of the order
of $\sim 30\ {\rm MHz}$. As a result, it is very unlikely that a
\textit{direct} observation of the induced EM emission would be
possible even from this simplified scenario.

Nevertheless, in the spirit of assessing whether this large release of
EM radiation can lead to \textit{indirect} observations of an EM
counterpart, it is useful to compare $E^{\rm rad}_{_{\rm EM}}$ with
the typical luminosity of an AGN. To fix the ideas let us consider
again a black hole of mass $M=10^8\msun \simeq 10^{41}\,{\rm g} \simeq
10^{61}\,{\rm erg}$, so that the luminosity in EM waves for
$B_0=10^4\,{\rm G}$ will be
\begin{eqnarray} 
\label{eq:lum}
L_{_{\rm EM}} \equiv \frac{E^{\rm rad}_{_{\rm EM}}}{\tau}  
&\simeq& 10^{41}\left(\frac{B}{10^4\ {\rm G}}\right)^2 {\rm
  erg\ s}^{-1} \nonumber \\
&\simeq& 10^{8}\left(\frac{B}{10^4\ {\rm G}}\right)^2 L_{\odot}\,,\nonumber \\
&\simeq& 10^{-4}\left(\frac{B}{10^4\ {\rm G}}\right)^2 L_{\rm Edd}\,,
\end{eqnarray}
where we have assumed a timescale $\tau \simeq 10^3\,M\simeq
10^5\,{\rm s}\simeq 1\,{\rm d}$ and where $L_{\odot}, L_{\rm Edd}$ are the
total luminosity of the Sun and the Eddington luminosity $L_{\rm Edd}
= 3.3\times 10^4\, (M/M_{\odot}) L_{\odot}$, respectively. While this
is a rather small luminosity (distant quasars are visible with
much larger luminosities of the order $10^{47}\, {\rm erg\ s}^{-1}$),
it is comparable with the luminosity of nearby AGNs and that is of the
order of $10^{41}\, {\rm erg\ s}^{-1}$. More important, however, is
the comparison between the EM emitted by the merging binary and the
one coming from the accretion disc. Using~\eqref{eq:lum} it is
straightforward to deduce that the binary EM luminosity is comparable
with that of an AGN accreting at $10^{-4}$ the Eddington rate. Hence,
unless the accretion rate is rather small (namely, much smaller than
$10^{-4}$ the Eddington rate with the extreme case being 
the non-accreting scenario) the EM emission from the binary would be
not only restricted to very low-frequencies but also just a small
fraction of the total luminosity. 
Under these conditions it is
unlikely that such emission could have an observable impact on the
overall luminosity of the accreting system.

As a final consideration it is useful to estimate whether the
inspiralling binary could nevertheless imprint a detectable effect on
the disc via the perturbations in the magnetic field it can
produce. To assess whether this is the case we first compare the
frequency $f_{_{\rm B}}$ with the typical plasma frequency
\begin{equation}
f_{_{\rm P}} = \frac{\omega_{_{\rm P}}}{2\pi} = 
\left(\frac{n_{e} e^2}{\pi m_e}\right)^{1/2} \simeq 10^{14} 
\left(\frac{n_{e}}{10^{21}\ {\rm cm}^{-3}}\right)\, \hz\,, 
\end{equation}
where $n_e$ is the electron number density, or with the electron
cyclotron $f_{_{\rm C}}$ frequency
\begin{equation}
f_{_{\rm C}} = \frac{\omega_{_{\rm C}}}{2\pi} = \frac{e B}{2\pi m_e c}
\simeq 10^{10} \left(\frac{B}{10^{4}\ {\rm G}}\right)\, \hz\,.
\end{equation}
Clearly, the magnetic field varies with a frequency $f_{_{\rm B}}$
that is is between $14$ and $18$ orders of magnitude smaller and hence
that the electrons and protons in the disc are always able to
``adjust'' themselves to the changes in the magnetic fields, which
are extremely slow when compared with the typical timescales in
the plasma. Stated differently, the EM radiation produced by the
inspiral cannot penetrate the disc and will be effectively reflected
over a skin depth of $\lambda = c/\omega_e \simeq 8\times
10^{-6}\ {\rm cm}$.

Finally, we consider whether the perturbed magnetic magnetic field can
have impact on the transport of angular momentum in the disc and hence
modify its accretion rate in a detectable way. It is worth remarking,
in fact, that there is considerably large EM energy flux reaching the
accretion disc and that is $F_{_{\rm EM}} \simeq L_{_{\rm
    EM}}/r^2_{\rm in} \sim 10^{11} (B/10^4\,{\rm G})^2\, {\rm
  erg\ s}^{-1}{\rm \ cm}^{-2}$, where $r_{\rm in}\sim 10^2\, r_g$ is
the inner radius of the disc and $r_g \simeq 10^{15}\ {\rm cm}$ is the
gravitational radius for a black hole of $10^8\ \msun$. A crude way to
estimate the perturbation on the disc is by considering the ratio
between the viscous transport timescale $\tau_V$ and the magnetic
transport timescale induced by the oscillating magnetic field,
$\tau_B$. Should this ratio be of the order of unity (or larger),
then the magnetic-field perturbation may be transmitted to the disc
in the form of Alfv\'en waves. In practice we estimate this by
considering the (inverse) ratio between the viscous and magnetic
torques, with the first one being expressed in terms of the average
pressure $p$ and sound speed $c_s$ as $f_{\phi, V} \simeq \alpha p
\simeq \alpha \rho c^2_s$ and the second one as $f_{\phi, B} \simeq r
\delta B^{\phi} B^z\alpha/(8 \pi) \simeq r \beta B^2_0/(8\pi)$; here
$\alpha$ is the standard alpha-disc viscosity parameter and $\beta$ is
a measure of the perturbation induced in the background magnetic field
(\ie $\delta B^{\phi} \sim \beta B_0,\ B^{z} \sim B_0$). We therefore
obtain
\begin{eqnarray}
\label{eq:taus}
\frac{\tau_V}{\tau_B}&=&\frac{f_{\phi, B}}{f_{\phi, V}} 
\\ \nonumber
&\simeq&
10 \frac{\beta}{\alpha}
\left( \frac{r}{10^{-2}\ r_g} \right)
\left( \frac{10^{-2} {\rm g\ cm}^{-3}}{\rho} \right)
\left( \frac{B_0}{10^4\ {\rm G}} \right)^2
\left( \frac{c}{c_s} \right)^2\,,
\\ \nonumber
\end{eqnarray}
where $10^{-2}\ r_g$ is the typical length scale over which magnetic
torques could operate. Assuming now $\alpha \simeq 0.1-0.01$, $\beta
\sim 10^{-2}$ and $c_s ~ 0.1-0.01\, c$ as reference numbers, the rough
estimate~\eqref{eq:taus} suggests that it is indeed possible that
${\tau_V} > {\tau_B}$ and hence that the perturbations in the magnetic
field, albeit small and rather slow, can induce a change in the
viscous torque and hence induce a change in the accretion rate if the
latter is sufficiently stable. Determining more precisely whether this
modulation in the magnetic field can effectively leave an imprint on
the accretion flow would require a more accurate modeling of the
accretion disc and is clearly beyond the scope of this simple
estimate. It is however interesting that this possibility is not
obviously excluded.

In summary, the analysis carried out in this Section shows that it is
highly unlikely that the EM emission associated with the scenario
considered in this paper can be detected \textit{directly} and
simultaneously with the GW one. This is essentially because the EM is
too inefficient for realistic values of the magnetic fields and
because it operates at frequencies which are well outside the ones
accessible to astronomical radio observations. However, if the
accretion rate of the circumbinary disc is sufficiently stable over
the timescale of the final inspiral and merger of the black-hole
binary, then it may be possible that the EM emission will be
observable \textit{indirectly} as it will alter the accretion rate
through the magnetic torques exerted by the distorted magnetic field
lines. A firmer conclusion of whether this can actually happen in
practice will inevitably have to rely on a more realistic description
of the accretion process.

As a final comment we stress that our analysis and discussions have
not included the role of gas or plasmas around the black hole(s) nor
have we considered resistive scenarios. Both of these ingredients,
when coupled to the EM fields behavior described here, could induce
powerful emissions by accelerating charged particles via the strong
fields produced (\eg in a manner similar to the Blandford-Znajek
mechanism~\cite{Blandford1977}) or by affecting the gas/plasmas
dynamics or via the reconnection of the complex EM fields produced
during the inspiral and merger. Future work in these directions is
needed in order to shed light on these possibilities and assess their
realistic impact as EM counterparts to the GW emission.

%
%
\section{Concluding Remarks}
\label{sec:conclusions}

We have analyzed the phenomenology that accompanies the inspiral and
merger of black-hole binaries in a uniform magnetic field which is
assumed to be anchored to a distant circumbinary disc. Our attention
has been concentrated on binaries with equal masses and equal spins
which are either aligned or antialigned with the orbital angular
momentum; in the case of supermassive black holes, these
configurations are indeed expected to be the most common
ones~\cite{Bogdanovic:2007hp,Dotti:2009vz}.  Furthermore, this choice
allows us to disentangle possible precession effects and concentrate
on the EM fields dynamics as affected by the orbital motion of the
binary.  Overall, the simulations reveal several interesting aspects in the
problem:
\begin{itemize}
\item The orbital motion of the black holes distorts the essentially
  uniform magnetic fields around the black holes and induces a
  quadrupolar electric field analogous to the one produced by the Hall
  effect for two conductors rotating in a uniform magnetic field. In
  addition, both electric and magnetic fields lines are dragged by the
  orbital dynamics of the binary. As a result, a time variability is
  induced in the EM fields, which is clearly correlated with the
  orbital behavior and ultimately with the GW-emission. The EM fields
  become, therefore, faithful tracers of the spacetime evolution.

\item As a result of the binary inspiral and merger, a net flux of
  electromagnetic energy is induced which, for the $\ell=2,m\pm2$
  modes is intimately tied, via a constant scaling factor in
  amplitude, to the gravitational energy released in GWs. This
  specular behaviour in the amplitude evolution disappears for
  higher-order modes, even though the phase evolution remains the same
  for all modes.

\item Because the tight correlation between the EM and the GW-emission
  has been found for all of the cases considered here, we expect it to
  extend to all possible binary configurations as long as the EM fields are
  playing the role of ``test fields''. Hence, the modelling of the GW
  emission does in practice provide information also on the EM one
  within the scenario considered here.

\item Although the global \textit{large-scale} structure of the EM
  fields is dictated by the orbital motion, the individual spins of
  the black holes further distort the EM field lines in their
  vicinities. These \textit{small-scale} fields may lead to
  interesting dynamics and to the extraction of energy via
  acceleration of particles along open magnetic field lines or via
  magnetic reconnection.

\item The energy emission in EM waves scales quadratically with the
  total spin and is given by $E^{\rm rad}_{_{\rm EM}}/M \simeq
  10^{-15} \left({M}/{10^8\ M_{\odot}}\right)^2 \left({B}/{10^4\ {\rm
      G}}\right)^2$, thus being $13$ orders of magnitude smaller than
  the gravitational energy for realistic magnetic fields. This EM
  emission is at frequencies of $\sim 10^{-4}(10^8 M_{\odot}/M)\,\hz$,
  which are well outside those accessible to astronomical radio
  observations. As a result, it is unlikely that the EM emission
  discussed here can be detected \textit{directly} and simultaneously
  with the GW one.

\item Processes driven by the changes in the EM fields could however
  yield observable events. In particular we argue that if the
  accretion rate of the circumbinary disc is small and sufficiently
  stable over the timescale of the final inspiral, then the EM
  emission may be observable \textit{indirectly} as it will alter the
  accretion rate through the magnetic torques exerted by the distorted
  magnetic field lines.

\end{itemize}

All of these results indicate that the interplay of strong
gravitational and EM fields represents a fertile ground for the
development of interesting phenomena. Although our analysis is
incomplete as the effects on plasmas are not taken into account, we
believe that the main properties of the EM dynamics described above
should hold as long as the energy in the black holes dominates the
energy budget. A more precise estimate of the possible emissions and
of the observational signatures calls for further studies which would
necessarily have to include additional physics. This work, however,
together with those in
refs.~\cite{2008ApJ...672...83M,Palenzuela:2009yr,chang,vanMeter:2009gu,Corrales:2009nv,Palenzuela:2009hx,Bode:2009mt},
constitute interesting first steps in this direction.

%
%
\acknowledgments

\noindent It is a pleasure to thank Kyriaki Dionysopoulou, Olindo
Zanotti, Chris Thompson, Avery Broderick, Steve Liebling and David
Neilsen for useful discussions and comments. DP has been supported as
VESF fellows of the European Gravitational Observatory
(EGO). Additional support comes from the DFG grant SFB/Transregio~7
``Gravitational Wave Astronomy'', NSF grant PHY-0803629 and NSERC
through a Discovery Grant. Research at Perimeter Institute is
supported through Industry Canada and by the Province of Ontario
through the Ministry of Research \& Innovation.  The computations were
performed at the AEI, the LONI network (\texttt{www.loni.org}), LRZ
Munich, and Teragrid.

%
%

\bibliography{draft,aeireferences}

%
%
\end{document}